\documentclass[final,letterpaper,2pt,times,twocolumn]{elsarticle}

\journal{TexExchange}

    \usepackage[letterpaper, portrait, margin=0.50in]{geometry}

    \usepackage{hhline}  
    \usepackage{tgtermes}
    \usepackage{tikz}

\usepackage{subfig}
\usepackage{geometry}
\usepackage{graphicx}
\usepackage{mathtools}
\usepackage{booktabs}
\usepackage{multicol}
\usepackage{multirow}
\usepackage{threeparttable}
\usepackage{float}
\usepackage{epstopdf}
\usepackage{rotating}
\usepackage{xcolor}
\usepackage{ragged2e}

    \fboxsep=\tabcolsep\relax
    \fboxrule=\arrayrulewidth\relax

\biboptions{sort&compress,comma,square}

\newcommand{\etal}{\textit{et al}.}

\def\etal{\mbox{\it et al.\ }}

\newcommand{\eq}[1]{Eq.~\ref{#1}}

\newcommand{\fig}[1]{Fig.~\ref{#1}}

\newcommand{\tab}[1]{Table~\ref{#1}}

\begin{document}


\begin{frontmatter}

\title{Probing discontinuous precipitation in U-Nb}

\author[msen]{Thien Duong}
\author[lanl]{Robert E. Hackenberg}
\author[msen]{Vahid Attari}
\author[llnl]{Alex Landa}
\author[llnl]{Patrice E.A. Turchi}
\author[msen,meen]{Raymundo Arr\'{o}yave\corref{cor1}}


\cortext[cor1]{\textit{Email address:} rarroyave@tamu.edu (Raymundo Arr\'{o}yave)}

\address[msen]{Department of Materials Science and Engineering, Texas A$\&$M University, College Station, TX 77843-3123, United States}

\address[meen]{Department of Mechanical Engineering, Texas A$\&$M University, College Station, TX 77843-3123, United States}

\address[lanl]{Los Alamos National Laboratory, P.O. Box 1663, Los Alamos, NM 87545, United States}

\address[llnl]{Lawrence Livermore National Laboratory, 7000 East Ave., Livermore, CA 94550-9234, United States}

\begin{abstract}

U-Nb's discontinuous precipitation, $\gamma^{bcc}_{matrix} \rightarrow \alpha^{orth}_{cellular} + \gamma'^{bcc}_{cellular}$, is intriguing in the sense that it allows formation and growth of the metastable $\gamma'$ phase during the course of its occurrence. Previous attempts to explain the thermodynamic origin of U-Nb's discontinuous precipitation hypothesized that the energy of $\alpha$ forms an intermediate common tangent with the first potential of the double-well energy of $\gamma$ at the $\gamma'$ composition. While this hypothesis is eligible and consistent with the experimental observation of gradual increase in $\gamma'$ composition at increasing temperature, it is in conflict with recent experiments whose results indicated a distribution of $\gamma'$ compositions in the vicinity of 50 at\%Nb. To shed some light onto this issue, the current work investigates the origin of U-Nb's discontinuous precipitation in view of fundamental thermodynamics and kinetics, taken from the perspective of phase-field theory. It has been showed that local misfit strain tends to play a crucial role in the formation and growth the discontinuous precipitation. Depending on the magnitude of strain developed at grain boundaries, either an increasing $\gamma'$ composition or a random distribution of $\gamma'$ composition around the equiatomic value with respect to increasing temperature could be expected.

\end{abstract}

\begin{keyword}
phase-field modeling \sep metallic fuels \sep U-Nb \sep thermodynamics \sep discontinuous precipitation
\end{keyword}

\end{frontmatter}



\section{Introduction}

    \fontdimen2\font=0.5ex
    \justify
Given its high melting point, good corrosion resistance, good conductivity and continuous bcc region at high temperatures, U-Nb is considered a promising candidate for Gen-IV fast breeder reactor. The material, however, exhibits various metastable phase transformations whose resulting microstructures strongly affect the fuel's performances (see, for instance, \cite{1980Vandermeer, 1984Eckelmeyer, 2007Volz, 2009Clarke}). In the current work, the phase transformation of interest is discontinuous precipitation (DP) \cite{1981Williams, 2001Manna} whose lamellar precipitate is known to degrade U-Nb's corrosion resistance and ductility \cite{2011Hackenberg}.

\begin{figure*}[ht!]
	\centering
	\subfloat[]{\includegraphics[width=0.99\columnwidth]{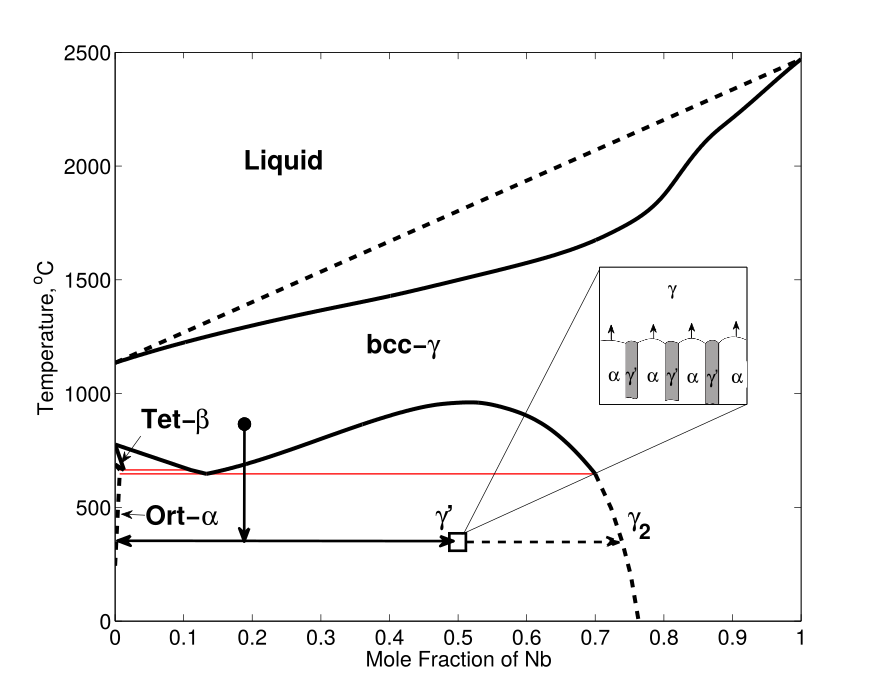}}
	\subfloat[]{\includegraphics[width=0.99\columnwidth]{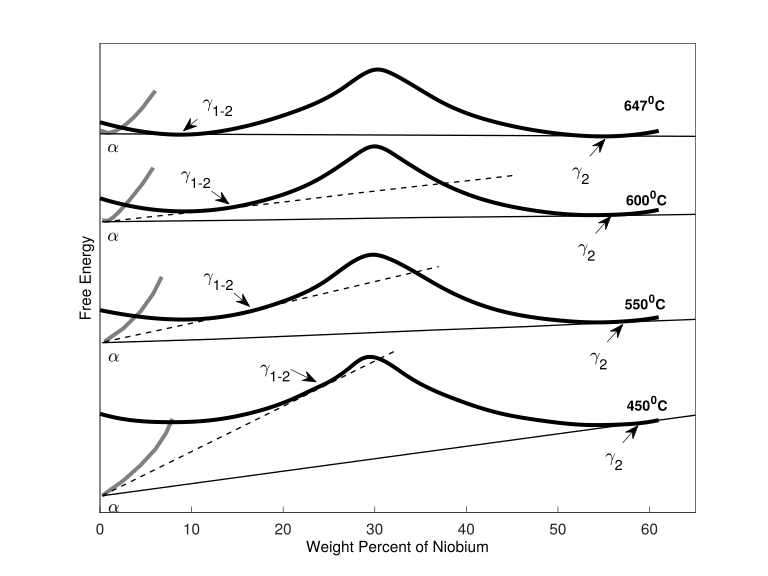}}
	\caption[]{(a) Schematic demonstration of discontinuous monotectoid decomposition in uranium - niobium system; (b) Schematic energies describing Djuric's hypothesis.}
	\label{fig:KoikeDjuricSchema}
\end{figure*}

DP is the result of a decomposition from a supersaturated solid solution into a solute-depleted matrix and a precipitate across moving boundary \cite{1981Williams, 2001Manna}. In the U-Nb system, DP occurs as part of the monotectoid decomposition:

\begin{equation*}
\gamma \xrightarrow{DP} \alpha + \gamma' \xrightarrow{DC} \alpha + \gamma_{2}
\end{equation*}
in which, $\gamma$ (or interchangeably $\gamma_{1}$) is quenched bcc matrix, $\alpha$ is orthorhombic precipitate, and $\gamma'$ (or interchangeably $\gamma_{1-2}$) is metastable bcc precipitate with an intermediate composition differing from that of stable $\gamma_{2}$ (see \fig{fig:KoikeDjuricSchema} (a)), DP is discontinuous precipitation and DC is discontinuous coarsening.

Although the observations of DP in the U-Nb system have been commonly reported, the thermodynamic and/or kinetic origin of its occurrence was scarcely addressed. Djuric \cite{1972Djuric} was perhaps the first and only who had provided an explicit explanation. In their work, Djuric examine the decomposition of the $\gamma$ phase in U-21.2Nb alloy. The studied samples are first homogenized at $950^\circ C$ for one week followed by water quenches. They are then solutionized at $900^\circ C$ for 24 hours and transfered to a tin bath for isothermal heat treatments at temperatures between $450^\circ C$ and $600^\circ C$. For longer isothermal heat treatments, double-stage vacuum furnace where the upper part is held at solution treatment temperature and the lower at isothermal treatment temperature is used. Based on the post-XRD analyses of the samples quenched from isothermal heat treatment, Djuric hypothesized that $\alpha$ and $\gamma$ form two local equilibriums (LE) with each other, one at an intermediate composition, $\gamma'$, and the other at the global equilibrium composition, $\gamma_2$. Due to the former LE, $\gamma$ would decompose partially into $\alpha$ and metastable $\gamma'$, if its initial composition $\gamma_1$ was less than $\gamma'$. This explains the occurrence of DP. If by chance $\gamma_2$ nucleated inside the system, the stable phase would set a lower energetic reference towards which the system would evolve spontaneously, resulting in discontinuous coarsening (DC). To demonstrate their hypothesis, Djuric schematically described energy profiles whose replications are shown in \fig{fig:KoikeDjuricSchema} (b).


It should be noted that Djuric's hypothesis is established upon their observation, i.e. the gradual increase of $\gamma'$ composition at increasing temperature. While the hypothesis is thermodynamically eligible, Djuric's observation is, interestingly enough, found contradicting with the recent experiments from Hackenberg \etal \cite{2011Hackenberg}. In their study, Hackenberg \etal \cite{2011Hackenberg} examine DP and DC of U-13Nb and U-17Nb (at.\%). Here, although the alloy compositions are different, the same phase transformations are observed, albeit the sizes of the precipitates can be slightly different. The samples are homogenized for 6 hours at $1000^\circ C$, solutionized for 30 minutes at $800^\circ C$ (U-13Nb) or $850^\circ C$ (U-17Nb), and water quenched before being annealed isothermally at various temperatures between $300^\circ C$ and $500^\circ C$. Via light optical and scanning electron microscopies, the authors observe DPs with $\gamma'$ compositions in the vicinity of 50 at.\%Nb. Insight into the thermodynamic driving force of DP as well as those of general precipitation (before DP) and DC (after DP) are then given by tracking the energy path across the reactions, towards equilibrium \cite{2015Hackenberg}. Although this information mediates the understanding why specific length scales, growth rates, and phase compositions are selected, it does not infer explanations for the occurrence and growth of $\gamma'$ during DP, especially with the phase's composition residing near the center of miscibility gap's unstable region (see \cite{1998Koike, 2008Liu, 2016Duong}). What it does instead is to challenge the validity of Djuric's experiments and hypothesis as, Djuric's observations (in term of $\gamma'$ compositions) are phenomenologically different from those of Hackenberg \etal's and as such their hypothesis can not be applied to explain Hackenberg \etal's results. This essentially rises intriguing questions about the origin of the discontinuous reaction and how it can be correlated to different experimental observations. 

To shed some light towards the addressing of these questions, we attempt, in the current work, the investigations of DP's origin from the fundamental thermodynamic and kinetic points of view. We choose to take our views from the interesting perspective of phase-field theory that accounts both thermodynamic and kinetic factors towards microstructural evolutions \cite{2002Chen, 2008Moelans, 2009Steinbach, attari2016phase, attari2018interfacial}. The proposed phase-field framework for the current analyses is the phase-field model with finite interface, or in short interface dissipation model, recently developed by Zhang and Steinbach \etal \cite{2012Steinbach, 2012Zhang}. Comparing with previous phase-field \cite{2009Amirouche, 2011Amirouche} and sharp-interface \cite{1955Turnbull, 1959Cahn, 1972Fournelle, 1991Fournelle, 1972Hillert, 1982Hillert, 1973Sundquist, 1997Klinger, 2001Purdy, 2013Robson} models, the interface dissipation model offers the uniqueness for assessing the thermodynamic evolution of the kinetic process at the moving interface. This can help to shed light into the nature of the discontinuous reaction. Following are our detailed analyses starting with Djuric's hypothesis.


\section{Thermodynamic backgrounds}

\begin{figure}[ht!]
	\begin{center}
		\includegraphics[width=0.50\columnwidth]{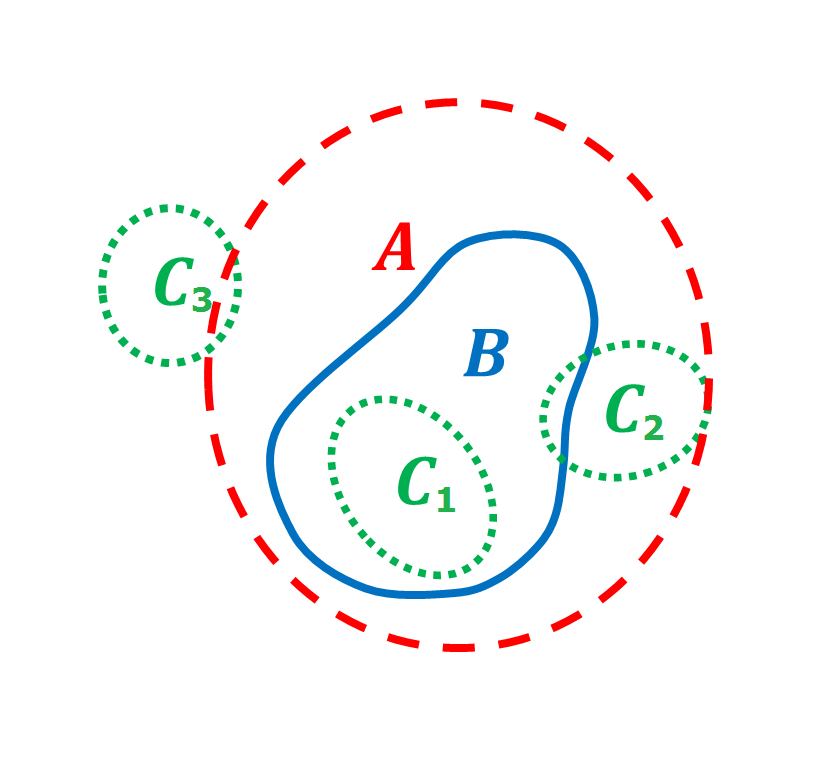}
		\caption[Set theory to demonstrate the usage of CALPHAD thermodynamic reference in examining Djuric's hypothesis]{Set theory to demonstrate the usage of CALPHAD thermodynamic reference in examining Djuric's hypothesis. Here, A represents the thermodynamics of U-Nb by nature; B represents the ordinary thermodynamics of U-Nb approximated by CALPHAD; C represents the two-LE phenomenon hypothesized by Djuric \cite{1972Djuric}.}
		\label{fig:Set}
	\end{center}
\end{figure}

To examine Djuric's hypothesis on the origin of U-Nb's discontinuous precipitation, the interface dissipation model was implemented in place of a free-energy minimizer to investigate the CALPHAD free energies of $\alpha$ and $\gamma$ for their possible LE (states of minimal energy). The idea is to utilize the verified CALPHAD description as a quantitative reference to examine Djuric's phenomenological hypothesis. The fundamental significance of this reference lies in the fact that it allows the thermodynamic hypothesis to be checked against the (by-nature) thermodynamics of U-Nb. To better demonstrate this, we adapted the simple set theory as shown in \fig{fig:Set}.

Within this figure, (set) $A$, $B$, and $C$ indicate the by-nature thermodynamics of U-Nb, the ordinary thermodynamics of U-Nb approximated by CALPHAD, and Djuric's phenomenological hypothesis, respectively. Here, the exact boundary of $A$ is unknown hence it is represented by a dashed line; $B$ is however known explicitly and believed to belong to $A$ since the CALPHAD assessment is in reasonable agreement with experiments \cite{2016Duong} (note that $B$ is believed to represent the core of $A$ - the ordinary thermodynamics defining U-Nb's chemical equilibria); therefore, it is proposed to be an available thermodynamic reference to examine $C$. According to the set theory, $C$ has three possible positions relative to $B$:

\begin{itemize}

\item $C_{1}$ belongs to $B$, i.e. Djuric's two-LE hypothesis is an inherent feature of the CALPHAD ordinary thermodynamics. Since $B$ also belongs to $A$, this makes $C_{1}$ is the subset of $A$. In other words, unless there exists a stronger mechanism to stabilize the metastable $\gamma'$ phase, it is highly possible that the thermodynamic hypothesis explains the discontinuous monotectoid decomposition.

\item $C_{2}$ intersects $B$, i.e. the two-LE hypothesis is only partially accounted for by the CALPHAD description. Although this is not as conclusive as in the first case, it still makes $C_{2}$ one possible subset of $A$. In other words, the thermodynamic hypothesis may still be valid. In this case, further investigations are needed to see if $C_{2}$ could possibly belong to A.

\item $C_{3}$ is far away from $B$, i.e. the CALPHAD description does not feature two LE at all. This makes $C_{3}$ less likely to be a subset of $A$. As a consequence, the possibility of Djuric's hypothesis being valid, is low. Here, it would also require further investigations as in the second case to verify whether $C_{3}$ really does not have any thermodynamic correlations with $A$.

\end{itemize}

\section{Computational details}

\begin{figure}[ht!]
	\begin{center}
		\includegraphics[width=0.80\columnwidth]{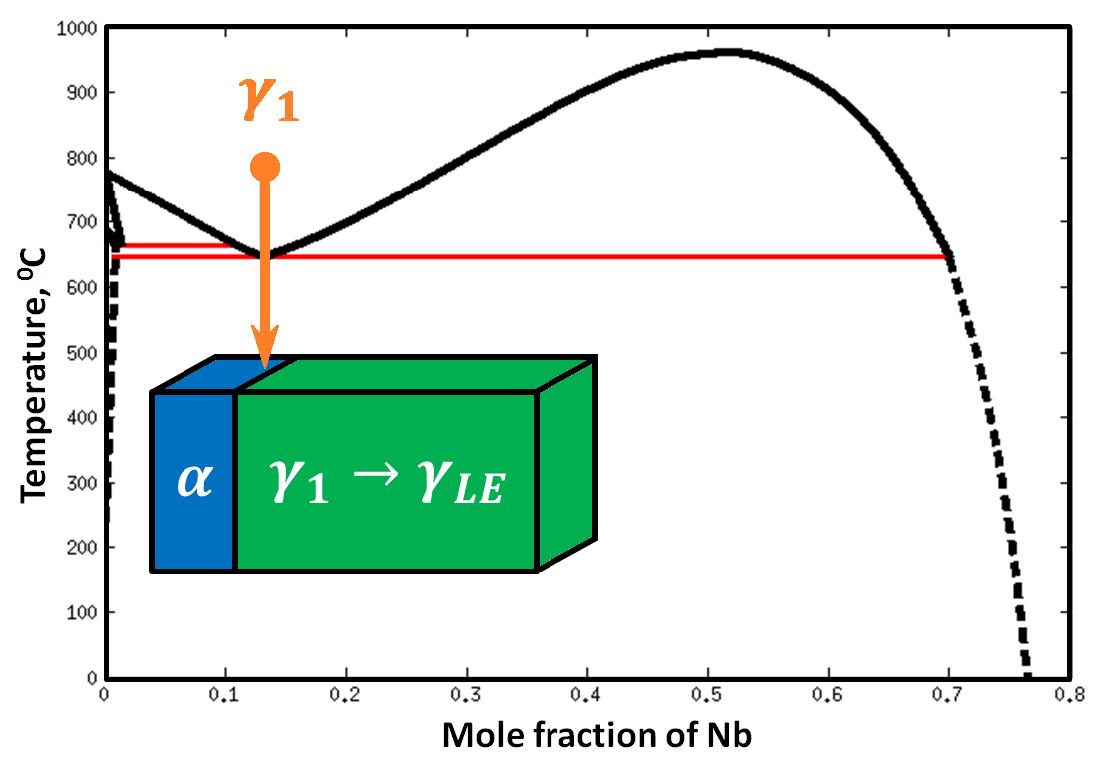}
		\caption[Schematic representations of diffusion-couple simulations to investigate the LE between $\alpha$ and $\gamma$]{Schematic representations of diffusion-couple simulations to investigate the LE between $\alpha$ and $\gamma$.}
		\label{fig:SchemDiffCouple}
	\end{center}
\end{figure}

To find possible LE between the CALPHAD energies of $\alpha$ and $\gamma$ using the proposed phase-field model, it is found that the diffusion-couple-type simulation is ideal for its simplicity, i.e. 1-D, yet sufficient, i.e. LE is sufficiently indicated when the Kirkendall interface stops moving, its driving force vanishes, and the compositions of two reacting phases are homogeneous. The schematic demonstration of the phase-field diffusion couple is shown in \fig{fig:SchemDiffCouple}.

The couple of interest is $1$ $\mu m$ long and initially consists of $0.1$ $\mu m$ of $\alpha$ and $0.9$ $\mu m$ of $\gamma$. It has a total of $500$ grid points with the step of $2$ $nm$ and interfacial width of $10$ $nm$. The initial compositions of $\alpha$ and $\gamma$ are 1 $at.\%$ Nb and 13 $at.\%$ Nb respectively.

To estimate thermodynamic driving force (in form of energy density [$J/cm^3$]), the previously assessed CALPHAD energetic data are used \cite{2016Duong}; for simplicity, the used molar volumes for this estimation are assumed to be constant and take the average value of those from the initial orth and bcc phases. The interfacial energies are taken as the averages of those evaluated in \cite{2011Hackenberg}.

The interfacial mobilities are chosen according to our empirical formula: $\mu = s\frac{D_{i}M_{i}}{D_{i} + M_{i}}$, where $s=10^6$ is a scaling factor, $i$ indicates either $alpha$ or $\gamma$, $M$ is the value of atomic mobility, and $D$ is the value of interdiffusivity. Here, the atomic mobility and interdiffusivity of $\gamma$ are taken from the previously assessed DICTRA database \cite{2016Duong}; basing on the atomic packing factors of orth and bcc, it was assumed that the atomic mobility and diffusivity of $\alpha$ are three times faster than those of $\gamma$; since these kinetic coefficients do not affect the thermodynamic LE between the two reacting phases, their precise values are of only peripheral interest within the scope of this work. Nevertheless, it is noted that kinetic factors can play an important role in determining the lamellar microstructure of DP, as demonstrated later in this work; therefore a comprehensive knowledge of these physical quantities is beneficial for future developments and applications of the nuclear material.

The summary of model parameters and physical parameters used in this work is given in \tab{tab:PFMParams}. For solving the model's evolution equations, finite-difference method was utilized. The numerical stability of this linear solver was supported by dynamic time step. 

\begin{table}[t]
  \centering
  \resizebox{\columnwidth}{!}{%
  \begin{threeparttable}[b]
    \caption[Numerical and material parameters for the diffusion-couple simulations]{Numerical and material parameters for the diffusion-couple simulations.}
    \begin{tabular}{l@{\hskip 36pt}l@{\hskip 36pt}l}
    \toprule
    Parameters & Symbols & Values \\
    \midrule
    Grid spacing & $\Delta x$   & $2.0$ $nm$ (1D), $1.5$ $nm$ (2D) \\
    Molar Volume & $V_M$ & $12.27$ $cm^3/mol$ \tnote{\textbf{a}} \\
    Interface energy & $\sigma_{\alpha\gamma'}$ & $0.14 \times 10^{-4}$ $J/cm^2$ \tnote{\textbf{b}} \\
                     & $\sigma_{\alpha\gamma}$  & $0.35 \times 10^{-4}$ $J/cm^2$ \\
                     & $\sigma_{\gamma'\gamma}$ & $0.31 \times 10^{-4}$ $J/cm^2$ \\                                      
    Permeability & $P_{\alpha\gamma}$ & $\frac{8M_\gamma}{a\eta}$ \\
    Lattice Parameter & $a$     & $3.5$ $\AA$ \tnote{\textbf{c}}\\
    Interface width & $\eta$     & $10.0$ $nm$ \\
    Atomic mobility of $\alpha$ & $M_\alpha$     & $3\times M_{\gamma}$ \\
    Atomic mobility of $\gamma$ & $M_\gamma$     & Database \cite{2016Duong} \\
    Diffusivity of $\alpha$ & $D_{\alpha}$ & $3\times D_{\gamma}$ \\
    Diffusivity of $\gamma$ & $D_{\gamma}$ & Database \cite{2016Duong} \\
    Interface mobility & $\mu_{\alpha\gamma}$ & $\frac{\varsigma D_{\gamma} M_{\gamma}}{D_{\gamma}+M_{\gamma}}$ $cm^4/Js$ \tnote{\textbf{d}}\\
    \bottomrule
    \end{tabular}%
      \begin{tablenotes}
        \item[\textbf{a}] \textit{Approximate average of $\alpha$-U and $\gamma$-U-50 at. \% Nb molar volumes (taken from the EMTO data)}           
        \item[\textbf{b}] \textit{Approximate averages of $\sigma_{\alpha\gamma}^{DP}$ and $\sigma_{\alpha\gamma}^{DC}$ reported in \cite{2011Hackenberg}}
        \item[\textbf{c}] \textit{Effective lattice parameter of choice, corresponding to $V_M$}     
        \item[\textbf{d}] \textit{Proposed empirical formula where $\varsigma=10^7$}
      \end{tablenotes}
    \label{tab:PFMParams}%
  \end{threeparttable} %
  }
\end{table}%

\section{Local equilibria}

\begin{figure*}[ht!]
  \begin{center}
      \subfloat[]{\includegraphics[width=0.90\columnwidth]{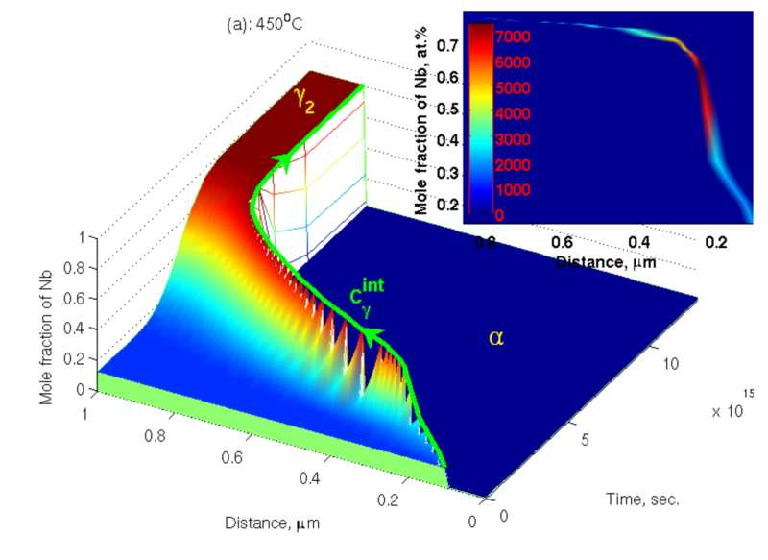}}
      \subfloat[]{\includegraphics[width=0.90\columnwidth]{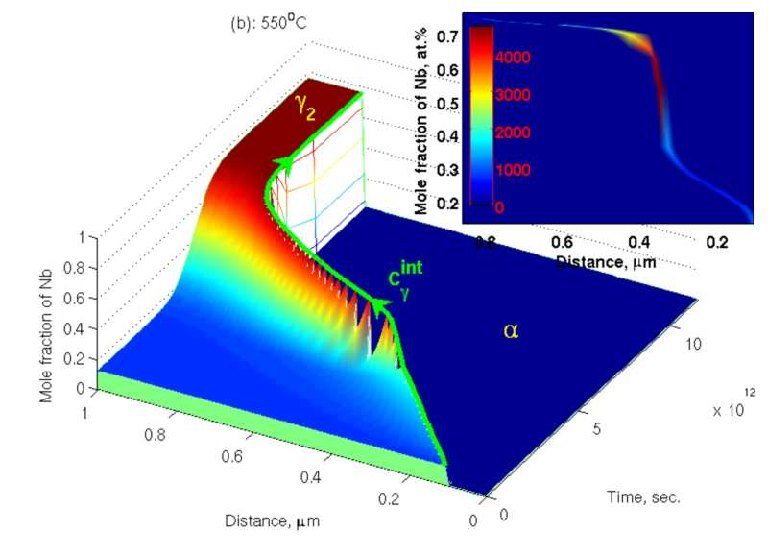}} \\
      \subfloat[]{\includegraphics[width=0.90\columnwidth]{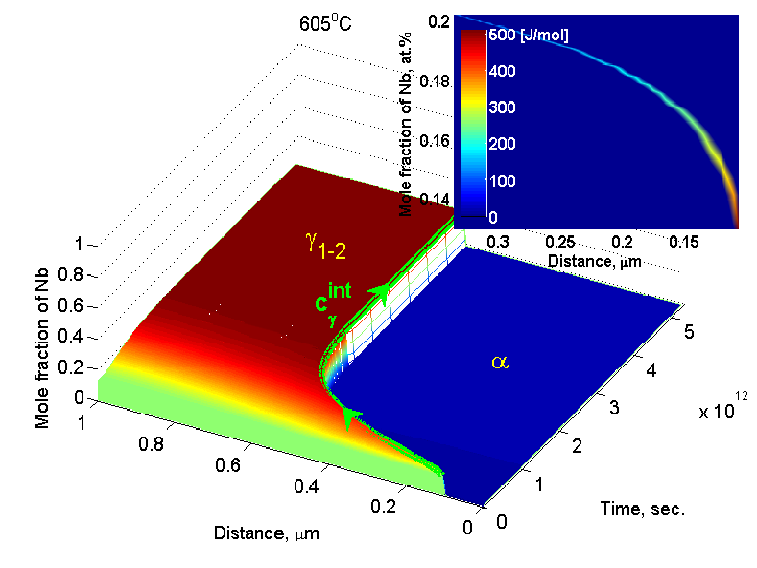}}
      \subfloat[]{\includegraphics[width=0.90\columnwidth]{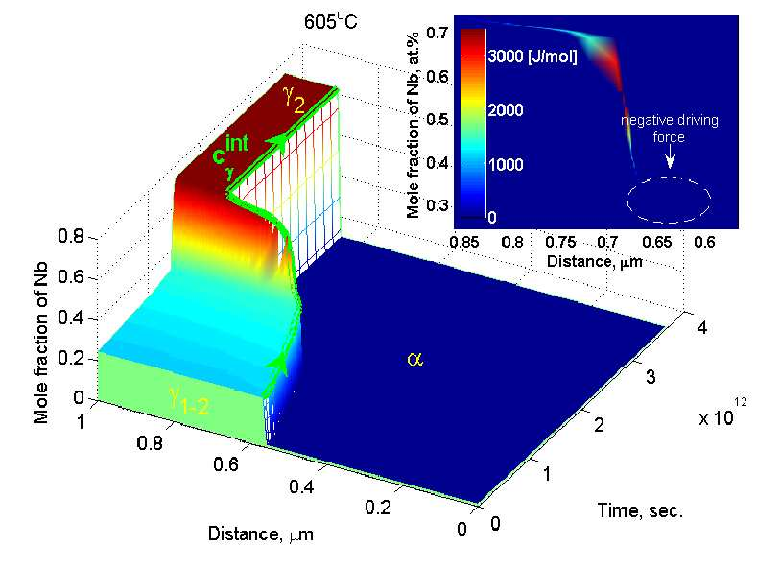}}
      \caption[Phase-field investigations of possible LE between $\alpha$ and $\gamma$ at (a) $450 ^\circ C$, (b) $550 ^\circ C$, and at (c, d) $605 ^\circ C$]{Phase-field investigations of possible LE between $\alpha$ and $\gamma$ at (a) $450 ^\circ C$, (b) $550 ^\circ C$, and at (c, d) $605 ^\circ C$. Here, the $\alpha$-growing / $\gamma$-shrinking processes of diffusion couples occur from right to left of the figures; the insets are the projections of the 3-D evolutionary paths of $c^{int}_{\gamma}$ on the `Mole fraction of Nb at.\%' - `Distance' plane and their colors indicate the magnitudes of the average chemical driving forces, $\bigtriangleup g^{phi}_{\alpha\beta}$, at the interface along these paths.}
      \label{fig:PFMDiffusionCouple}
  \end{center}
\end{figure*}

Simulation results of the phase-field diffusion couples at $450^\circ C$ and $550^\circ C$ are shown in \fig{fig:PFMDiffusionCouple} (a) \& (b). Here, the 3-D plots represent the evolution of the diffused $\alpha|\gamma$ interface with respect to spatial distance (x-axis), time (y-axis) and composition (z-axis). The solid (green) lines with arrows in the 3-D plots indicate the evolutionary path (and directions) of the composition of $\gamma$ at the interface ($c^{int}_{\gamma}$) during the phase transformation. The insets feature the average chemical driving force, $\bigtriangleup g^{phi}_{\alpha\beta}$, at the diffusion-couple interface plotted as a function of $c^{int}_{\gamma}$ and the position of the interface (for details about $\bigtriangleup g^{phi}_{\alpha\beta}$ please check \cite{2012Steinbach}). The color within the insets shows the order of magnitude of the chemical driving force (note that non-zero driving force is only located around the projection of the evolutionary path of $c^{int}_{\gamma}$ on the distance -- composition (x--z)  plane and that since the initial setups of the diffusion-couple simulations are out-of-equilibrium all average chemical driving forces at the beginnings of the simulations are non-zero).

It can be seen from \fig{fig:PFMDiffusionCouple} that after the evolution time is larger than $1.0\times10^{14}$ ($s$) for $450^\circ C$ or $2.0\times10^{11}$ ($s$) for $550^\circ C$ the Kirkendall interface stops moving; the interface's chemical driving force, $\bigtriangleup g^{phi}_{\alpha\beta}$, vanishes (see insets); and, the compositions in both $\alpha$ and $\gamma$ reach their homogeneous values across the phase regions. These all indicate that the interface dissipation model has found, for each temperature, one LE, at which the $\gamma$ composition is identical to that of the stable $\gamma_2$ ($79$ at.\% Nb for $450^\circ C$ and $76$ at.\% Nb for $550^\circ C$ as in \cite{2016Duong}). Some notes are:

\begin{itemize}

\item The sluggish evolution time (of orders $10^{14}$ ($s$) for $450^\circ C$ and $10^{11}$ ($s$) for $550^\circ C$) results from the estimated CALPHAD's slow bulk diffusivity (in orders of $10^{-23}$ $[cm^2/s]$ for $450^\circ C$ and $10^{-21}$ $[cm^2/s]$ for $550^\circ C$ \cite{2016Duong}, consistent with the experimental values from Peterson and Ogilvie \cite{1960Peterson, 1963Petersonb}). In reality, the reaction happens much faster due to the fast boundary-diffusion condition at the reaction front of DP \cite{2001Manna} as evidenced by the measured interphase boundary diffusivity triple products in \cite{2011Hackenberg}. 

\item The sudden increase in $\bigtriangleup g^{phi}_{\alpha\beta}$ from $40$ to $60$ at.\% Nb for both simulation cases (red areas in insets) corresponds to the period within which $c^{int}_{\gamma}$ evolves through the center of the unstable region of the bcc miscibility gap. The significant driving force within this region urges $c^{int}_{\gamma}$ to quickly leave the unstable region for the following low-energy area of the $\gamma_2$ LE, creating essentially two noticeable necking points: one around $30$ at.\% Nb (vicinity of the first inflection point) and the other around $60$ at.\% Nb (vicinity of the second inflection point), along the evolutionary path of $c^{int}_{\gamma}$ for both simulation cases.

\end{itemize}

\noindent Further phase-field investigations of $\alpha-\gamma$ LE for both $450^\circ C$ and $550^\circ C$ with initial compositions of $\gamma$ higher than $\gamma_2$ all showed a convergence back to the same equilibrium state at $\gamma_2$. This essentially indicates that within the CALPHAD energy landscape at $450^\circ C$ and $550^\circ C$ the orthorhombic phase, $\alpha$, only form with the bcc phase, $\gamma$, one LE which corresponds to the stable $\alpha+\gamma_2$ products of the monotectoid decomposition; no LE can be found at the intermediate composition of $\gamma'$ as hypothesized by Djuric. As a matter of fact, it is found that single LE, i.e. $\alpha+\gamma_2$, is a common phenomenon throughout the temperature interval between $400^\circ C$ and $600^\circ C$ (actually up to $605^\circ C$), within which Djuric's experiments were carried out.

Given that the CALPHAD energies \cite{2016Duong} should be better references than those phenomenologically hypothesized by Djuric \cite{1972Djuric}, would this defy the two-local-equilibrium hypothesis?

Interestingly enough, we found that, within the higher temperature range from $605^\circ C$ to $647^\circ C$, the CALPHAD energies do form two LE with each other and the first LE does lead to the thermodynamic state of DP, very much consistent with Djuric's hypothesis. The simulation results of the phase-field diffusion couple at $605^\circ C$ are shown in \fig{fig:PFMDiffusionCouple} (c) \& (d). As evidenced by this figure, the interface dissipation model finds two LEs: one at an intermediate $\gamma'$ composition of $24.49$ at.\% Nb, as shown in \fig{fig:PFMDiffusionCouple} (c), and the other at the stable $\gamma_2$ composition of $73.95$ at.\% Nb, as shown in \fig{fig:PFMDiffusionCouple} (d). Note especially in \fig{fig:PFMDiffusionCouple} (d) that there is an interesting region within which the chemical driving force is negative and external driving force has to be artificially introduced to compensate for the negative value and to get the diffusion-couple system to evolve; the significance of this negative driving force is elucidated as follows.


Within the framework of phase-field modeling, the process of finding the LE at $605^\circ C$ progressed as follows:

\begin{itemize}
\item First, phase-field diffusion couple was started with the initial composition of $\gamma$ at $13$ at.\% Nb. After the evolution time was greater than $2\times10^{11}$ ($s$), it was observed that the Kirkendall interface stopped moving, $\bigtriangleup g^{phi}_{\alpha\beta}$ converged to a value of almost 0 $J/mol$, and the compositions within the bulk phases reached their homogeneous states. In other words, the interface dissipation model found the system's first (or intermediate) LE whose $\gamma$'s composition was $24.49$ at.\% Nb. Note here that the $\bigtriangleup g^{phi}_{\alpha\beta}$ distribution along the evolutionary path of $c^{int}_{\gamma}$ and correspondingly the morphology of the path (inset of \fig{fig:PFMDiffusionCouple} (c)) are different from those at lower temperatures (see insets of \fig{fig:PFMDiffusionCouple} (a) and (b)); in particular, they do not exhibit unusual peak (red in color) and necking points along the evolution process respectively; this is due to the fact that $c^{int}_{\gamma}$ has not yet passed through the first inflection point of the bcc miscibility gap to enter the gap's unstable region. Further prolonging the simulation did not lead to any significant changes. The almost-zero $\bigtriangleup g^{phi}_{\alpha\beta}$ at the found LE utterly shut down the system and trapped it there.

\item In order to break this stasis and continue the phase-field investigation of $\alpha-\gamma$ LE, the composition of $\gamma$ was slightly shifted to a higher value while keeping the composition of $\alpha$ unchanged. The simulation result interestingly showed that, within this small deviation, the system tended to converge back to its initial LE. This was because here $\bigtriangleup g^{phi}_{\alpha\beta}$ had negative values which tended to reverse the compositional increment in order to bring down the system's total energy. In other words, there existed a finite energy barrier after the first LE which tentatively prevented the system from further evolving to higher $\gamma$-composition after the first LE. This energy barrier together with the vanishing driving force (as described above) form an effective two-fold obstacle which proceeds to interrupt the monotectoid decomposition and cause DP.

\begin{figure}[ht!]
  \begin{center}
 	  \includegraphics[width=0.90\columnwidth]{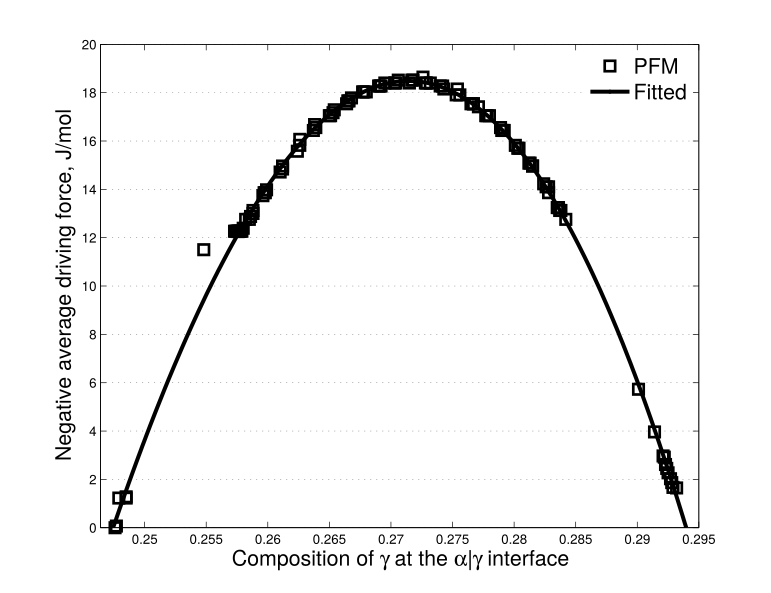}
      \caption[Energy barrier introduced by the intermediate local equilibrium after the $\gamma'$ composition to hinder the discontinuous monotectoid decomposition]{Energy barrier introduced by the intermediate local equilibrium after the $\gamma'$ composition to hinder the discontinuous monotectoid decomposition. Note that negative average driving force at the interface, $-\bigtriangleup g^{phi}_{\alpha\beta}$, is reported in this figure. }
      \label{fig:EnergyBarrier}
  \end{center}
\end{figure}

To force the system to overcome the energy barrier, a positive driving force was artificially introduced into the reacting interface in order to counter the negative value of $\bigtriangleup g^{phi}_{\alpha\beta}$ when it was observed. This artificial driving force could in reality be legitimated by the fact that the relaxation of internal stresses (due to volume/strain mismatch) between $\alpha$ and $\gamma$ lamellae (with $\gamma'$ composition) after some sufficient aging time, will essentially break down the first LE between $\alpha$ and $\gamma$ (by altering their free energies to lower values) and likely put the system into an out-of-equilibrium condition with non-trivial thermodynamic driving force to continue evolving in the DC manner \cite{1972Djuric, 2011Hackenberg}. During the introduction of artificial driving force, it was observed that the peak of the energy barrier that the system had to overcome was about $18.65$ $J/mol$, as shown in \fig{fig:EnergyBarrier}.

\item When $\bigtriangleup g^{phi}_{\alpha\beta}$ turned positive, the artificial driving force was removed to allow the evolution of the system to resume as normal. At this moment, the system had already entered the unstable region of the bcc miscibility gap. The driving force here was so significant that it dramatically drove the system almost instantaneously out of the unstable region (inset of \fig{fig:PFMDiffusionCouple} (d)). When the system's $c^{int}_{\gamma}$ passed through the second inflection point of the miscibility gap, $\bigtriangleup g^{phi}_{\alpha\beta}$ started converging and eventually brought the system to its second LE located at the $\gamma_{2}$ composition of $73.95$ at.\% Nb. Note here that the entire subprocess after the DP reaction (the first LE) practically represented the later DC reaction: $\alpha + \gamma' \rightarrow \alpha+\gamma_{2}$  \cite{1972Djuric, 2011Hackenberg}. After this, the system again stayed idle at the $\gamma_{2}$ LE. Further phase-field investigations at higher $\gamma$-compositions did not result in any additional LE. The interface dissipation model found a total of two LE between $\alpha$ and $\gamma$ in comparison to only one LE in the previous findings at lower temperatures.

\end{itemize}

\noindent To confirm this, an additional minimization was implemented in MATLAB to double check the number of LE between $\alpha$'s and $\gamma$'s CALPHAD free energies. The minimization is conventionally with respect to composition and at a specific temperature. The size of compositional domain for each LE search is controllable, and the considered temperatures are from both $400^\circ C - 605^\circ C$ and $605^\circ C - 647^\circ C$ ranges. It was found that the CALPHAD free energies of $\alpha$ and $\gamma$ indeed form two LE with each other within the temperature range of $605^\circ C - 647^\circ C$ while they exhibit only one LE within the temperature range of $400^\circ C - 605^\circ C$, consistent with the phase-field investigations.

According to the set theory proposition mentioned earlier (see \fig{fig:Set}), the observation of two LE between $\alpha$ and $\gamma$ within the temperature range of $605^\circ C-647^\circ C$ (or the union of $C$ and $B$) tentatively indicates that Djuric's hypothesis is a possible explanation for the origin of U-Nb's discontinuous monotectoid decomposition ($C$ possibly belongs to $A$, i.e. $C_{2}$). Yet, this indication is not conclusive due to the fact that the CALPHAD free energies show only one LE within $400^\circ C-605^\circ C$. To further investigate this, we revisited in the following the CALPHAD free energies of $\alpha$ and $\gamma$ within the temperature range between $400^\circ C$ and $605^\circ C$. 

\section{Strain effect }

\subsection{Strain energy}
As the first rather-\textit{ad hoc}-yet-simple attempt, we empirically sketched out new energetic profiles based on the CALPHAD free energies and following Djuric's proposal \cite{1972Djuric}. For this, the piecewise cubic polynomial with ten knots was used. This polynomial first allowed the accurate fittings of the CALPHAD base energies then enabled the desired modifications on top of these bases by fine tuning the positions of appropriate knots. The resulting energies are illustrated at $450 ^\circ C$ and $550 ^\circ C$ in \fig{fig:PFMEnergies1}. As demonstrated in this figure, the empirical estimations indicated that Djuric's hypothesis holds when the non-equilibrium energies around the lump of the bcc miscibility-gap is slightly or moderately increased. Since it is always the first impression that an increase in the total energy of a system is usually the result of strain/stress, these empirical findings lead us to the following considerations:

\begin{figure*}[ht!]
  \begin{center}
      \subfloat[]{\includegraphics[width=0.99\columnwidth]{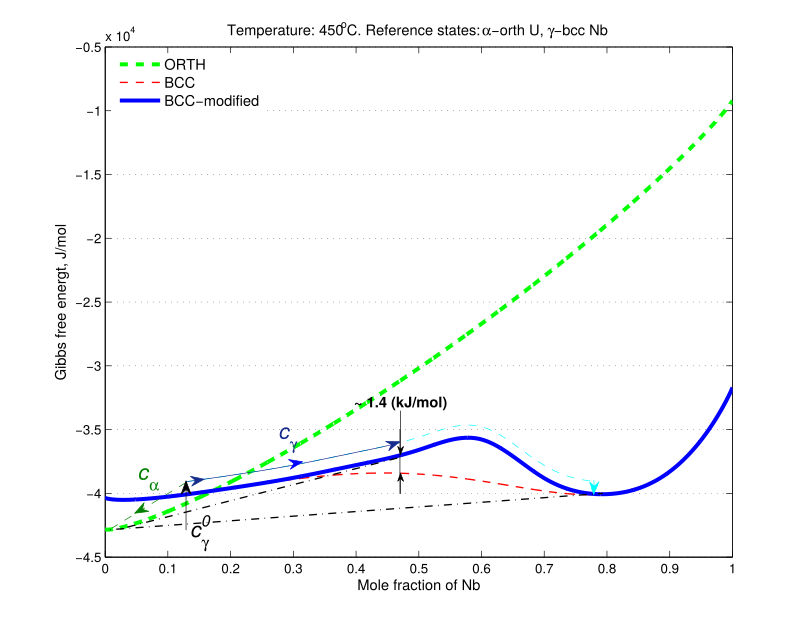}}
      \subfloat[]{\includegraphics[width=0.99\columnwidth]{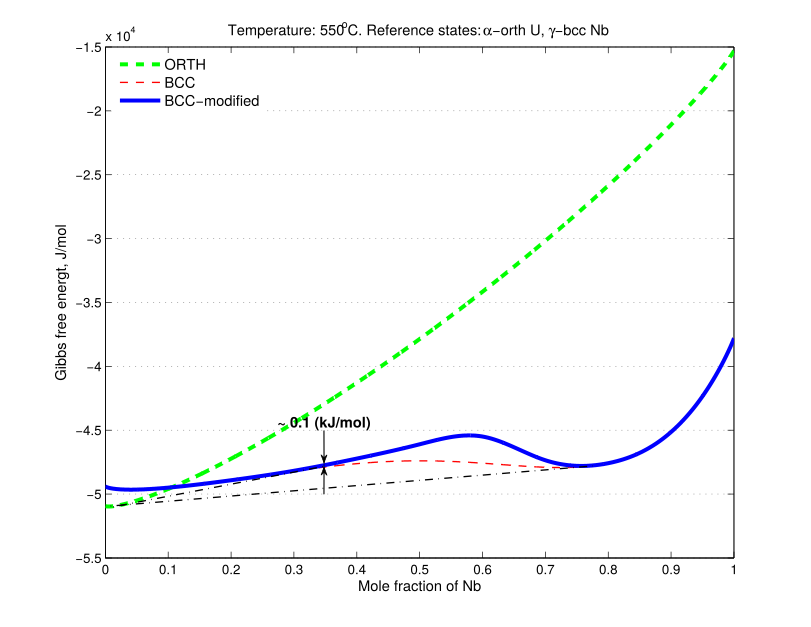}}
      \caption[Proposed strain-adjusted free energies of $\alpha$ and $\gamma$ at $450^\circ C$ (a) and $550 ^\circ C$ (b), plotted with reference to orth-U and bcc-Nb]{Proposed strain-adjusted free energies of $\alpha$ and $\gamma$ at $450^\circ C$ (a) and $550 ^\circ C$ (b), plotted with reference to orth-U and bcc-Nb. Here, since additional strain energy was assumed to be insignificant in $\alpha$, the phase's proposed free energies were chosen to be the same as CALPHAD free energies to simplify the effort. Notice that the proposed free energies form two common tangents with each other. These strain-adjusted free energies can be seen as the realization of Djuric's hypothetical free energies via CALPHAD methodology.}
      \label{fig:PFMEnergies1}
  \end{center}
\end{figure*}

\begin{itemize}

\item In the case of U-Nb's discontinuous monotectoid decomposition, due to the volume mismatch ($\sim 1\%-15\%$) between $\alpha$ ($20.8625 $ $\AA^3/Atom$ \cite{1963Eeles}) and $\gamma$ ($18.10$ (pure Nb) - $20.65$ (pure U) $\AA^3/Atom$ [This work]), there exists a stress/strain field at around the interfacial region between the two phases. This stress/strain field is distributed around the lamellae where discontinuous monotectoid decomposition happens, making its existence within the microstructure and effect on the decomposition non-ignorable.

\item CALPHAD free energies are bulk free energies. Although the bulk energies can be used to describe interfacial thermodynamics of the reaction front between $\alpha$ and $\gamma$ under ordinary condition where interfacial effects (e.g. coherency, stress/strain, etc.) are trivial, they tend to be insufficient under the opposite case. In such cases, additional energies raising from interfacial effects must be explicitly taken into account.

\item Since the energy raising from volumetric strain tends to be smaller at higher temperature (due to thermal relaxation), it is possible that the CALPHAD bulk energies can account for this energy under high temperature conditions, and hence the observations of two LE within $605^oC-647^oC$. In contrast, at lower temperature the (residual) strain energy tends to be larger and tends to deviate the system out of its ordinary thermodynamics, making the CALPHAD descriptions less sufficient. 

\end{itemize}

Combining the above considerations, it is believed that volumetric strain plays an important role in the stabilization of the intermediate $\gamma'$ phase during the discontinuous reaction. This stress/strain is supposed to distribute along the $\alpha|\gamma$ lamellar structures and its non-trivial associated energy deviates the system (locally) out of the ordinary thermodynamics represented by CALPHAD at temperatures ranging from $400^\circ C$ to $605^\circ C$ (and possibly lower). It is perhaps this elasto-chemical energy that legitimates Djuric's hypothesis, i.e. the union of C into A. To further investigate the possibility of this hypothesis, the elastic contribution of misfit strain is accounted for in an appropriated physical manner as follows:

\subsection{Harmonic approximation}

First, we adopt the general formula of:

\begin{equation}
f_{elas} = \frac{1}{2}\int_{v}\sum_{i,j}\sigma_{i,j}\epsilon_{i,j}dV
\label{eq:elasticenergy}
\end{equation}

While it is possible to augment, in a mathematically and thermodynamically consistent manner, the above formula into different phase-field models for quantitative descriptions and such an augmentation is not uncommon, it complicates a lot numerical implementations and often bury the thermodynamic big picture beneficial to the understanding of phase transformations beneath complicated mathematical operations. Here, to make it simple yet valid and relevant for flexible physical considerations, we make the following assumptions. First, we assume, since $\gamma$ is considerably softer than $\alpha$, that $\gamma$ is the only phase that is strained. Second, we assumed that the lattice parameter of the soft phase follows Vegard's law, albeit it was previously reported via first-principles calculations a small nonlinear behavior \cite{2016Duong}. We assumed further that there are no shear components at the interface. By applying the fourth-rank stiffness tensor to the linear elasticity \cite{2005Balluffi}, the elastic stresses read:

\begin{eqnarray}
\sigma_{zz} &=& 0 \\
\sigma_{xx,yy} &=& \frac{E}{1-\nu}\epsilon_{xx,yy} \\
\sigma_{xy} &=& \sigma_{yz} = \sigma_{zx} = 0
\end{eqnarray}

\noindent
where $\nu$ is the Possion's ratio and $E$ is the Young's modulus. It follows that \cite{cahn1961spinodal,yi2018strain,attari2019exploration}:

\begin{eqnarray}
f_{elas} &=& \frac{1}{2}\int_{v} \sum_{i,j}\sigma_{ij}\epsilon_{ij} dV \nonumber \\
         &=& \frac{E}{2(1-\nu)}\int_{v} \left( \epsilon^2_{xx} + \epsilon^2_{yy} \right)dV
\label{eq:reducedelasticenergy}
\end{eqnarray}

\subsection{Reliability range}

\begin{figure}[ht!]
	\centering
	\includegraphics[width=0.99\columnwidth]{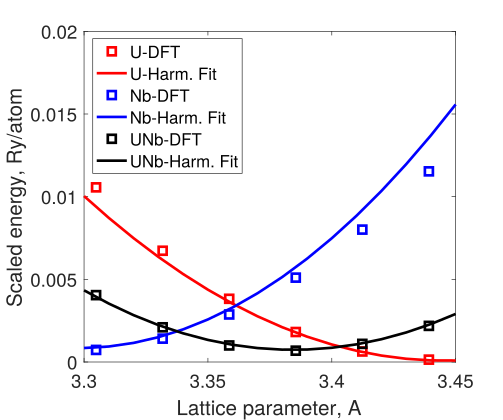}
	\caption[Validity of strain energy]{Validity of the harmonic approximation to the strain energy within the considered range of lattice parameter of $\gamma$.}    
	\label{fig:harmonic}
\end{figure}

Since the elastic energy is in the form of the harmonic approximation (second order), it is expected to underestimate/overestimate the anharmonic response of the system under high tensile/low compressive strain \cite{yi2018strain,attari2019exploration}. To identify the limit at which the harmonic approximation becomes less reliable, cohesive energetic response of the $\gamma$ phase was investigated using the Exact Muffin Tin Orbital method \cite{EMTO}. The results in which the first-principles anharmonic responses are compared with harmonic approximations are reported in \fig{fig:harmonic} at the compositions of $0$, $50$, and $100$ $at.\%$ Nb. Here, for simplicity the harmonic approximations are realized by using the data located within the harmonic proximity of the calculated cohesive curves. It is noted that the reported lattice parameters coincide with the entire atomic fraction of $\gamma$, i.e. from pure Nb on the left terminal to pure U on the right terminal of \fig{fig:harmonic}. As can be seen from this figure, harmonic approximation works best within the vicinity of 50 $at.\%$ Nb and tends to deviate from anharmonic behaviors near the end-members. Generally speaking, it is expected to be qualitatively acceptable within $\pm 30\%$ lattice deviation which is correspondent to $\pm 30$ $at.\%$ Nb deviation from which the lattice misfit between $\alpha$ and $\gamma$ is smallest.

It is also noted that the estimation of strain energy requires the identifications of habit planes between $\alpha$ and $\gamma$, which allows the estimation of misfit strain in \eq{eq:reducedelasticenergy}. In the occasion that lattice parameters change with changing composition during diffusion reaction, the identification of habit planes is composition dependent. This excludes the convenient adaptation of the $(111)_\gamma||(200)_\alpha$ habit plane of U-7.5Nb-2.5Zr \cite{1975Couterne} not only due to the alloy's fixed composition but also due to its different lattice parameters as compared to those of the binary. Since $\alpha$ is assumed earlier to not undergo misfit deformation and it does not deform by change of composition ($\alpha$'s composition appears almost constant \cite{2011Hackenberg}), this process reduces to a simpler case in which only the lattice parameter of $\gamma$ varies (during solute diffusion).

\subsection{Minimal-deformation plane}

    To identify the habit planes between $\alpha$ and $\gamma$ phases, we follow the principle of invariant line \cite{1982Dahmen} which assumes that the common line between matrix and precipitate lattices is the favored nucleation site of the precipitate. To avoid the cases in which the rotation required to match the common lines between precipitate and matrix lattices results in a large strain along the other direction that forms the matrix-precipitate interface with the common line, we require further that this direction is also an invariant line. In other words, an invariant plane as habit plane. Such invariant planes, however, rarely exists in practice. There exists instead common planes between matrix and precipitate with minimal lattice deformations. This results in a new approach, which we name minimal-deformation plane. This approach although being less (physically) constrained is more practical than the invariant-plane (or ideal common-line) approach. The proposed numerical algorithm to identify minimal-deformation planes is summarized in Table~\ref{tab:alghorithm}, and is as follows:


	First, super lattices are defined for the reacting structures. Here, we define the supercells of $2\times2\times2$ (2X), $3\times3\times3$ (3X), and $4\times4\times4$ (4X) for both $\alpha$ and $\gamma$ structures; and the cases of $(2X)_{\alpha}||(2X)_{\gamma}$, $(2X)_{\alpha}||(3X)_{\gamma}$, $(3X)_{\alpha}||(3X)_{\gamma}$, and $(4X)_{\alpha}||(4X)_{\gamma}$ are considered. The lattice parameters of $\alpha$ are collected from all literatures recorded in the ICSD database. For lowering computational expense, only the minimum, maximum, and mean values of the lattice parameters are considered for the evaluations of habit planes. The composition-dependent lattice parameters of $\gamma$ are adapted from Jackson's experiments \cite{1970Jackson}. Before selection, all parameters are converted to the same investigated temperature, hereinafter $450^\circ C$, using the thermal expansion coefficients taken from \cite{1956Bridge,2002Cverna}.
	
	
	Second, a triangle defining an interface within each super lattice is selected. Three edges of the triangle are estimated and compared to those of the other triangle in a sorted order. The pair of triangles with the lowest summation of squared edge mismatches is one correspondent to the habit planes. To avoid the case in which minimal misfit exists for the smallest cells containing the triangles but not repeatable throughout the interface, planar periodic condition is enforced for each considered triangle. This is done not based on the original (3D) lattice references -- which essentially gives rise to the need for explicit consideration of coupled translational and rotational degrees of freedom -- but on (2D) references defined by the triangles themselves -- which inherently imposes the translational and rotational matching throughout the interface.

    \newcommand{\tabitem}{~~\llap{\textbullet}~~}

    \begin{table*}[ht!]
    \caption{Algorithm I. Minimum Misfit Strain}
    \footnotesize
    \centering
    \begin{tabular}{l} \toprule
        \textbf{Algorithm I}: Minimal-deformation plane \\ \midrule
        1. \textbf{Define super-lattices/supercell:} lattice parameters and uncertainties chosen according to: \\
        \qquad Experimental (ICSD**) min, mean, and max values \\
        \qquad Temperature dependency accounted. \\
        2. \textbf{Find minimal-deformation plane} \\
             \qquad Establish 2 triangles, each defined arbitrarily by 3 atoms in corresponding super lattice (ie. establish lattice plane, see \fig{fig:HabitPlanes} for demonstration).\\
             \qquad Compare 3 edges of one triangle (bcc) to those of another (orth), ie. shortest to shortest and longest to longest.\\
             \qquad Estimate planar misfit as total edge-misfit (note: translational and rotational matching throughout the interface is inherent). \\
             \qquad Repeat until lowest planar mismatch\\
        3. \textbf{Estimate elastic energy:} according to edge mismatches.  \\
        ** ICSD stands as Inorganic Crystal Structure Database. \\ \bottomrule 
    \end{tabular}
    \label{tab:alghorithm}
    \end{table*}

\begin{figure*}[ht!]
	\begin{center}
\includegraphics[width=0.80\textwidth]{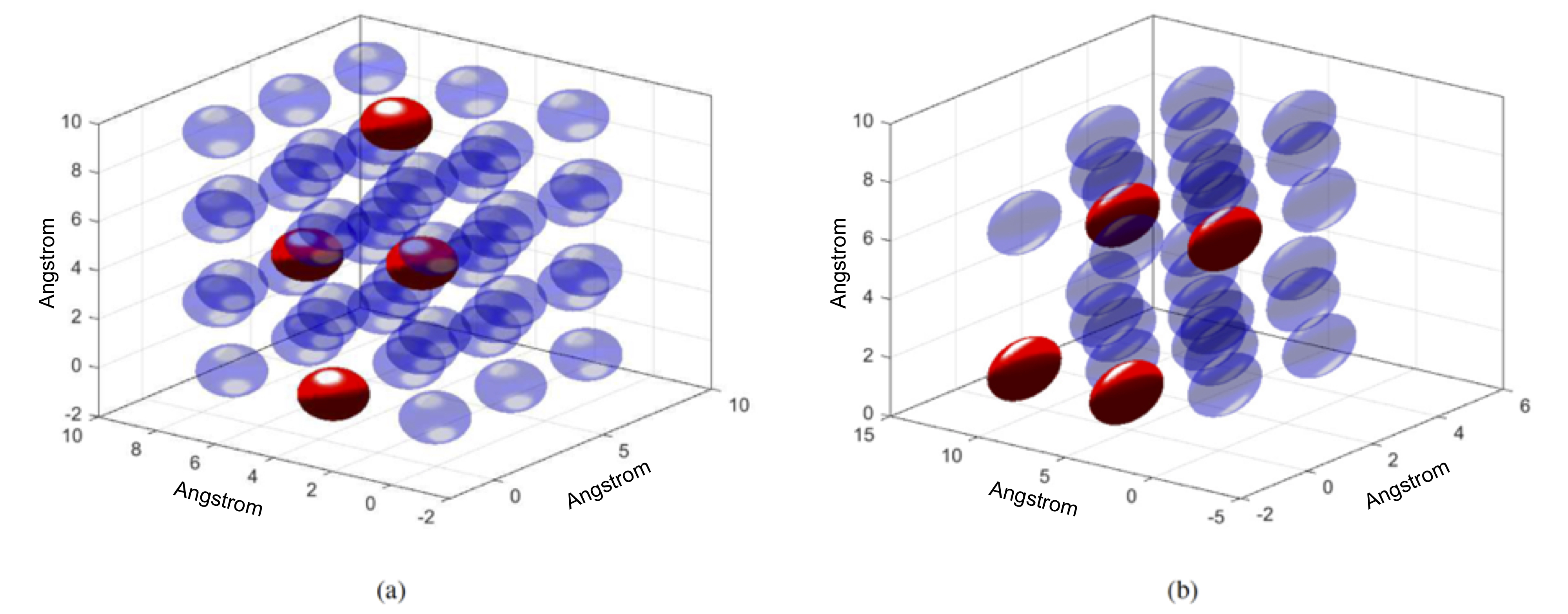}
		\caption[HabitPlanes]{A predicted (a)-$(3X)_{\gamma}$ $||$ (b)-$(2X)_{\alpha}$ habit plane. Of the four red atoms in each structure, three define the triangle. The last red atom is a periodic image of one of the triangle point, also highlighted for a better representation of the crystal plane.}
		\label{fig:HabitPlanes}
	\end{center}
\end{figure*}


\subsection{Stochastic elasto-chemical energies}
\begin{figure*}[ht!]
	\centering
	\subfloat[]{\includegraphics[width=0.90\columnwidth]{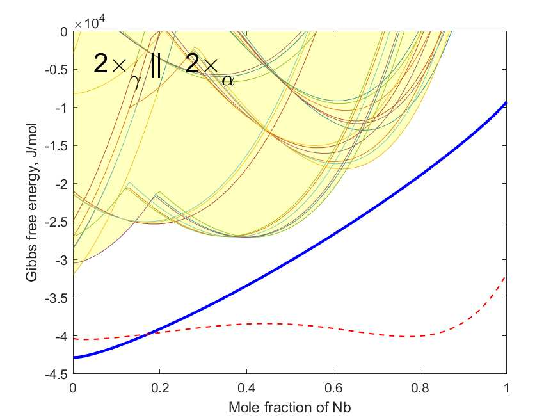}}
	\subfloat[]{\includegraphics[width=0.90\columnwidth]{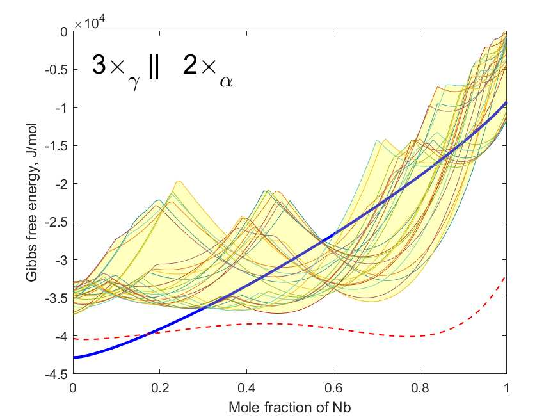}} \\
	\subfloat[]{\includegraphics[width=0.90\columnwidth]{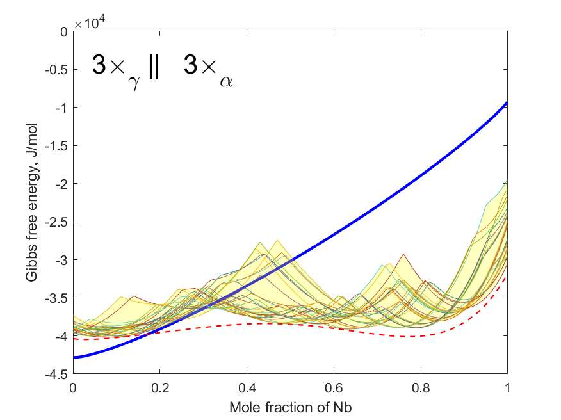}}
	\subfloat[]{\includegraphics[width=0.90\columnwidth]{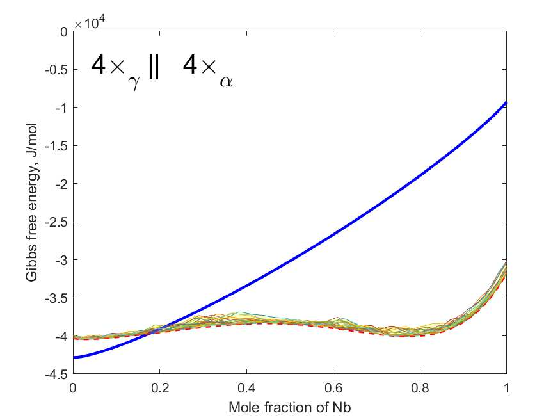}} \\
	\caption[Elasto-chemical energy]{Elasto-chemical energy of $\gamma$ assuming that the mechanically stronger $\alpha$ phase does not undergo any strain. Here, each elasto-chemical energy curve (e.g. the bold red curve) within the uncertainty band (i.e. filled area) composes of different local valleys. Each local valley corresponds to an estimated habit system that is lowest in strain energy. Each necking point along the curve represents the (1st order) transformation from one habit system to another (assuming that interface is a phase) when Nb content redistribute during DP. Note that the calculation of strain energy here only accounts for the misfit strain; and as such it favors habit system with larger (2-D) unit cells as these cells possess smaller misfit lattices; in practice, larger misfit unit cells would require higher formation energies, which could prevent a habit-plane structure from transformation to another structure along the diffusion path; estimation of such formation energy is unfortunately not simple.}    
	\label{fig:elastochemicalenergy}
\end{figure*}

The Young's moduli were adapted from Jackson's experiments \cite{1976Jackson}. The resulting stochastic elasto-chemical energies are reported in \fig{fig:elastochemicalenergy} for the cases of $(2X)_{\alpha}||(2X)_{\gamma}$, $(2X)_{\alpha}||(3X)_{\gamma}$, $(3X)_{\alpha}||(3X)_{\gamma}$, and $(4X)_{\alpha}||(4X)_{\gamma}$.

As can be seen from \fig{fig:elastochemicalenergy}, the estimated elastic energies are higher for smaller cell. For the case of $(2X)_{\alpha}||(2X)_{\gamma}$ the elasto-chemical energies are so high that even the stable $\gamma$ is hardly observable. $(2X)_{\alpha}||(2X)_{\gamma}$ is as such not the system's preference and is subjected to further transformation to other habit planes of lower energies. Within the current analysis, these habit planes are $(2X)_{\alpha}||(3X)_{\gamma}$, $(3X)_{\alpha}||(3X)_{\gamma}$, and $(4X)_{\alpha}||(4X)_{\gamma}$; the energy gain is considerably large for the former but not much so for the latter. Here, the reason for these lower energies is that the chance to find habit planes with smaller and smaller lattice misfit increases as more and more atoms are considered. These planes, however, are inevitably larger in matching units (i.e., the smallest misfit area between two periodic habit planes) and likely give rise to a larger barrier which the system has to overcome. Such barrier is not easy to assess; and for this reason, we choose $(2X)_{\alpha}||(3X)_{\gamma}$ (\fig{fig:elastochemicalenergy} (b)) and $(3X)_{\alpha}||(3X)_{\gamma}$ (\fig{fig:elastochemicalenergy} (c)) as, intuitively, the most likely habit planes among the predicted habit planes.

Along each elasto-chemical energy, there exists many energy valleys that are stabilized by local elastic strain. Of these energy valleys, many form common tangent with the $\alpha$ phase at intermediate compositions and as such can be subjected to DP. This indicates that (composition-dependent) elastic strain can affect the system's thermodynamic properties, non-trivially. Analogous to phase transformation, the inflection points along each energy correspond to (first-order) transitions from one set of habit planes to another set of habit planes. Since these points locate within the accepted bounds of $\pm$ $30$ $at.\%$ Nb, the used harmonic approximations are acceptable, according to the above analysis. It is noted here that the energies required for the interface transformations are relatively small, i.e. in order of a few $kJ/mol$ taking both elastic (see \fig{fig:elastochemicalenergy}) and interfacial energies (see \cite{2011Hackenberg}) into account, making the transformations competitive to dislocation formations\footnote{The energy required for forming a dislocation, $F \approx Gb/A$ where $G$ \cite{2013Jain} is shear modulus, $b$ is burger vector, and $A$ is molar area, is in the order of $10-100$ $kJ/mol \cdot atom$ (depending on the composition of $\gamma$) given the $<111>$ slip directions and two atoms reside along the burger vector.} which would otherwise arise due to the same need of relaxing excess strain




As evidenced by \fig{fig:elastochemicalenergy} (b) \& (c), there is a wide spectrum of elasto-chemical potentials, each results from one set of lattice parameters, and they all differs thermodynamically from each others. This emphasizes again that elastic strain plays an important role in the thermodynamic properties of the system.
Among the predicted potentials, it is noted that many curve do not appear to promote energy valleys that could be related to DP. These are believed to result from the overestimation of elastic energy due to the use of first-principles (0K) moduli. It is expected that these energies should be lower in practice and would as such show intermediate energy valleys that are more stable than $\alpha$. Such valleys increase the chance of having intermediate common tangents between $\alpha$ and $\gamma$, which in turn gives rise to the formation and growth of the intermediate $\gamma$ precipitate. 

\begin{figure}[ht!]
	\centering
	\includegraphics[width=0.99\columnwidth]{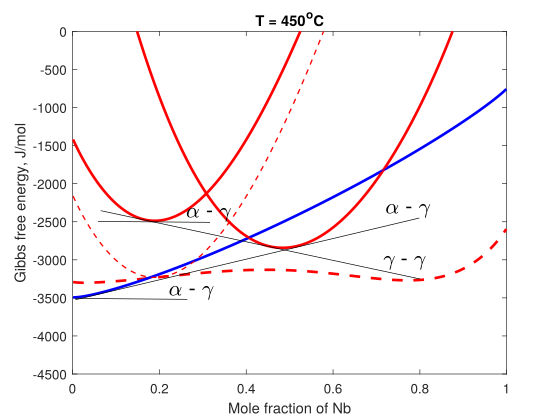}
	\caption[Energy diagram]{Energy diagram for the case where an elasto-chemical local energy valley resides within the vicinity of 50 at.\% Nb. This energy valley corresponds to the energy of the $\gamma$ precipitate. It forms (1) a common tangent with the free energy of the $\alpha$ precipitate and (2) a common tangent with $\gamma$ matrix's energy valley (which also belongs to the same elasto-chemical energy curve as that of the $\gamma$ precipitate). These two tangents define local equilibrium at the $\alpha$-$\gamma$ precipitate and $\gamma$ precipitate - $\gamma$ matrix respectively. At the interface between $\alpha$ precipitate and $\gamma$ matrix, the tangent can be either common tangent or parallel tangent, depending on the height of the matrix's energy valley. Together, these tangents govern DP's kinetic reaction.}    
	\label{fig:PFMEnergies2}
\end{figure}

The predicted potentials can be categorized into two phenomenological groups. 
The first group is characterized by two energy valleys: one locates near the equiatomic composition and the other near the composition of matrix. The potentials belong to this group are mainly distributed in \fig{fig:elastochemicalenergy} (a) and are the results of a large $\alpha$'s $a$ lattice parameter. Together with the energy of $\alpha$, the two characteristic valleys of this group form the typical sets of common/parallel tangents correspondent to a monotectoid decomposition. The thermodynamic and kinetic growth of DP in this case is as such expected to be similar to that of the eutectoid reaction. To further elucidate this, phase-field simulations are conducted assuming two common DP's kinetic conditions: volume-diffusion-controlled \cite{1959Korchynsky} and boundary-diffusion controlled \cite{2001Manna, 1981Williams}.

For the case of volume-diffusion-controlled DP, interfacial diffusivities are chosen to be equal to bulk diffusivities, taken from the recent assessment \cite{2016Duong}. For the case of boundary-diffusion-controlled, the bulk diffusivities are again taken from \cite{2016Duong}, the interfacial diffusivities at the reaction front/grain boundary are derived from the experimental interphase boundary diffusivity triple product $sD\delta$ \cite{2011Hackenberg}, where $s$ is the segregation factor at the interface, $D$ is the needed diffusivity, and $\delta=1$ (\AA) is interfacial width, and finally the diffusivities at $\alpha|\gamma'$ interface are chosen to be $10^3$ times smaller than those at the reaction front. The simulation domains are $192 \times 108$ $nm$ and $384 \times 108$ $nm$ in size, which reflects physical length scale of the system as measured in \cite{2011Hackenberg}. They initially consist of two $\alpha$ precipitates and two $\gamma$ precipitates whose compositions and sizes follows the LE partition. For simplicity, simplified energy diagram, as shown in \fig{fig:PFMEnergies2}, is used to represent thermodynamic driving force of this group. The energy values of the $\gamma$ phase are approximations to the local valleys at around 20 at.\% Nb and 45 at.\% Nb of the highlighted elasto-chemical energy in \fig{fig:elastochemicalenergy} (b).
	
\begin{figure}[ht!]
	\centering
	\includegraphics[width=0.80\columnwidth]{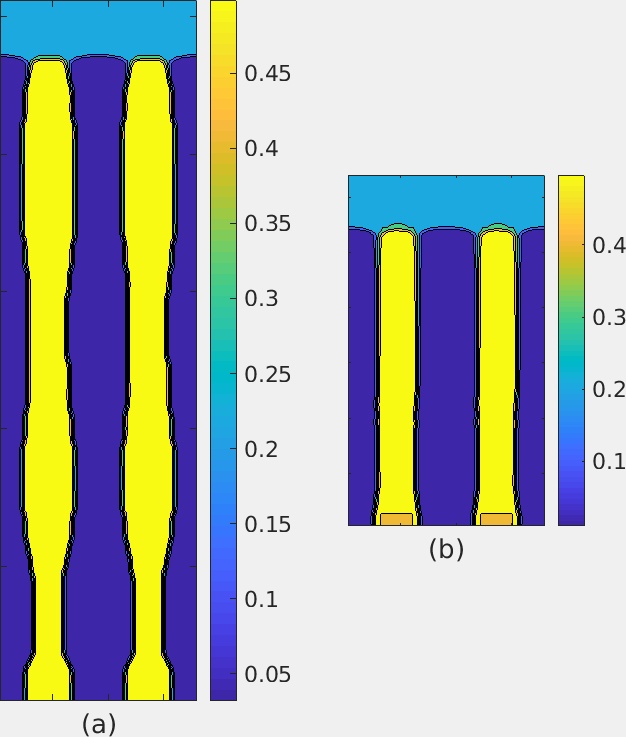}
	\caption[Phase-field simulation at $450^\circ C$ without (a) and with (b) fast boundary diffusion]{Phase-field simulation at $450^\circ C$ without (a) and with (b) fast boundary diffusion (note that color indicates the mole fraction of Nb). Domain size is a) 70$\times$255 $nm^2$ and b) 70$\times$130 $nm^2$. The bulk Nb is used as the matrix phase ($\gamma_1$) in the calculations.} 
	\label{fig:2DPFM_Hackenberg}
\end{figure}
	
The simulation result is reported in \fig{fig:2DPFM_Hackenberg}. As can be seen from the figure, stable growth of the intermediate $\gamma$ lamellae can be achieved in both cases. As similar to eutectoid decomposition, the reason for this is because the reaction is mainly governed by the system's thermodynamics, i.e. equilibrium is well defined at all interfaces of the reaction. The difference though is that instead of having static global equilibriums the system is trapped at local energy valleys along the dynamical decomposition from its initial matrix composition to final stable $\gamma_{2}$, i.e. similar to that demonstrated above in the 1-D phase-field analysis. Also, in the current phase-field analysis, the interface between $\alpha$ precipitate and $\gamma$ matrix is specially defined by a parallel tangent rather than the usual common tangent. This parallel tangent always leaves the interface out of equilibrium and as such there always exists a driving force to grow the $\alpha$ precipitate. The persistent force applies as well at the triple junction between the two precipitates and matrix and therefore would also support the growth of the $\gamma$ precipitate, essentially adding DP's overall growth rate. In the typical case when three common tangents exist between the phases -- which may well be missed due to existing uncertainty and used approximation, the growth would happen in a much more stable manner but at a slower rate. The role of kinetics in the stabilization of the metastable $\gamma$ phase as well as its (thermodynamically) stable growth is rather trivial. It, however, does affect the growth rate of the reaction and the morphologies of $\alpha$ and $\gamma$ phases, e.g. as shown in \fig{fig:2DPFM_Hackenberg}. Given as well that the incubation time for the nucleation of the stable $\gamma$ is longer than that of the intermediate $\gamma$ phase (due to its higher composition), the nucleation and (stable) growth of the intermediate $\gamma$ is expected from this (local equilibrium) thermodynamic and kinetic point of view. It is interesting to note here that, due to the special topological of the potentials within the first group (Potential defined by the two energy valleys), the estimated composition of the intermediate $\gamma$ is always found close to the equiatomic composition under different temperatures. This is found to be in good agreement with the recent experiments from Hackenberg \etal \cite{2011Hackenberg}.
	
Unlike the previous group, a second group also exists that features an energetic valley far below the equiatomic composition. Here after we refer to this category as the second group. The energy valley in the second group can either accommodate both the matrix and intermediate $\gamma$ compositions as in \fig{fig:elastochemicalenergy} (c) or only the composition of the precipitate leaving the lower matrix composition at a another energy valley as in \fig{fig:elastochemicalenergy} (a) (note that the $\alpha$'s $a$ lattice parameter is smaller here than in the first group). As temperature increases, such an energy valley will promote a gradual increase in the intermediate $\gamma$ composition as the common tangent between this precipitate and the $\alpha$ precipitate will be stretch to the higher Nb content. This is found to support the previous experimental observation and hypothesis of Djuric \cite{1972Djuric}. For the case when the matrix and precipitate compositions reside in two different energy valleys below the equiatomic composition, the thermodynamics and kinetics of the reaction are similar to those analyzed above. For the cases when the potentials have to accommodate the compositions of both matrix and intermediate $\gamma$ precipitate phases within one energy valley, the DP reaction is not so well thermodynamically defined as in the previous case. 

\begin{figure}[ht!]
	\centering
    \includegraphics[width=0.80\columnwidth]{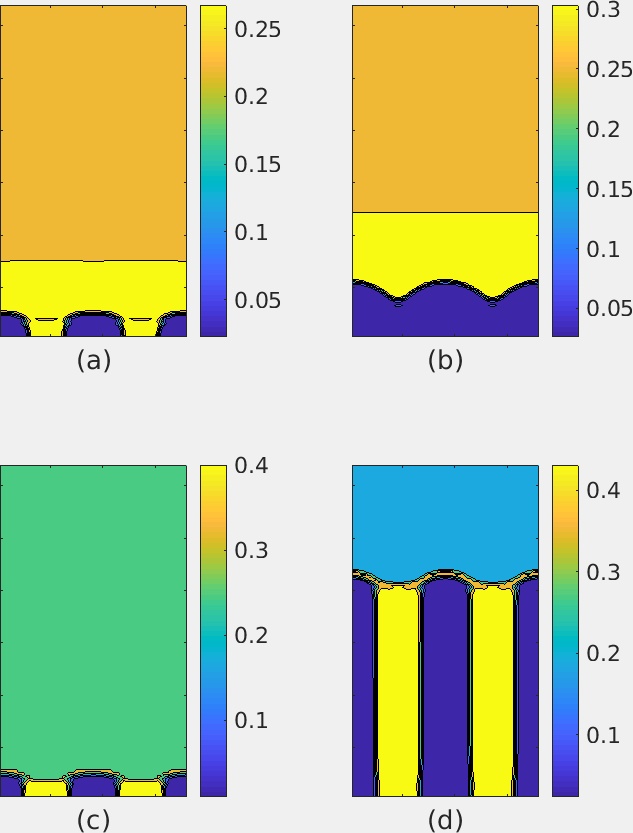}
	
	\caption[Phase-field simulation at $450^\circ C$ without (a, b) and with (c, d) fast boundary diffusion]{Phase-field simulation  at $450^\circ C$ without (a, b) and with (c, d) fast boundary diffusion (note that color indicates the mole fraction of Nb). Domain size is 70$\times$130 $nm^2$. The bulk Nb is used as the matrix phase ($\gamma_1$) in the calculations.}    
	\label{fig:2DPFM_Djuric}
\end{figure}

To investigate further, the same kinetic analyses as in the above case were conducted. The thermodynamic driving force is now qualitatively described by the energy diagram shown in \fig{fig:PFMEnergies1} (a). The results for the case of volume-diffusion-controlled is shown in \fig{fig:2DPFM_Djuric} (a) \& (b). As can be seen from these figures, the initially nucleated $\alpha$ precipitates eventually impinge and coalesce while the $\gamma_{1-2}$ precipitates fade out. The reason for this is that at the interface between the $\gamma_{1}$ matrix and $\gamma_{1-2}$ lamellae there occurs a down-hill diffusion between the two bcc phases, i.e. a Nb flux flows from $\gamma_{1-2}$ to $\gamma_{1}$. This flux (vertical flux) dissipates a considerably large amount of Nb content out of the $\gamma_{1-2}$ lamellae. Note that relative to this flux, there exists another flux (lateral flux) that flows along the tips of the $\alpha$ lamellae (due to the curvatures/gradients of these lamellae along the reaction front) and, in the opposite way, adds more Nb content to the $\gamma_{1-2}$ lamellae to grow them. Unfortunately, in this case the later lateral flux is slower than the vertical flux and not able to sustain the Nb content within the $\gamma$ lamellae near its LE value. This essentially breaks down the equilibrium between the $\alpha$ and $\gamma_{1-2}$ lamellae, allowing the $\alpha$ lamellae to expand into the $\gamma_{1-2}$ lamellae until impingement. The evolving system therefore does not exhibit the discontinuous reaction, and in this case Djuric's (thermodynamic) hypothesis falls short as being the only necessary and sufficient condition for the origin of the discontinuous reaction in the uranium-niobium system.
	
\begin{figure}[ht!]
	\centering
	\includegraphics[width=0.99\columnwidth]{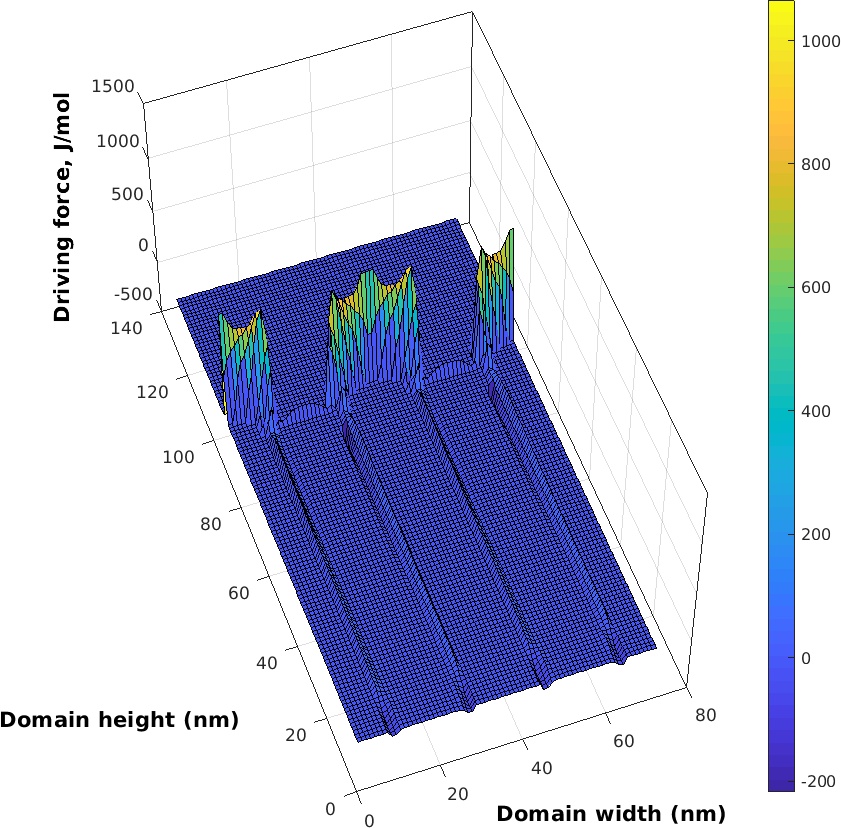}
	\caption[]{Driving force distributed along the interfaces for the boundary-diffusion-controlled case in group 2 Group with an energetic valley far below the equiatomic composition). The bulk Nb is used as the matrix phase ($\gamma_1$) in the calculations.}    
	\label{fig:2DPFM_Djuric_Force}
\end{figure}
	
For the case of boundary-diffusion-controlled, simulation result is reported in \fig{fig:2DPFM_Djuric} (c, d). Here, in contrast to the previous case, when the fast grain-boundary condition was taken into account, the Nb-flux flowing into the $\gamma_{1-2}$ lamellae from the tips of the $\alpha$ lamellae is much more significant than the Nb-flux flowing out of the $\gamma$ lamellae due to the down-hill diffusion. In other words, there is not much Nb leakage from the $\gamma$ lamellae into the $\gamma$ matrix and its LE state with the $\alpha$ lamellae is sustained during the reaction. Analysis of driving force at the $\alpha|\gamma_{1-2}$ interfaces shown in \fig{fig:2DPFM_Djuric_Force} demonstrates this. In this figure, it is also observed that the driving force distributed at the $\gamma_{1-2}|\gamma_{1}$ reaction front is considerably smaller than that at the $\alpha|\gamma_{1}$ interface. The existence of this driving force during down-hill diffusion is found to be in good agreement with Hillert's theory \cite{1972Hillert, 1982Hillert}, which states that the driving energy for the growth of $\gamma_{1-2}$ grain is identified with some fraction of the free energy which \textit{``would be lost due to volume diffusion if certain mechanisms did not interfere}'' \cite{1972Hillert}. In our case, there exists such interfering mechanisms, i.e. the fast flux along the reaction front acts as resistance to the down-hill diffusion, and hence the non-trivial driving force at the $\gamma_{1-2}|\gamma_{1}$ interface. Yet, it appears that the effective growth of the $\gamma_{1-2}|\gamma_{1}$ reaction front is due more to the carrying role of the growing $\alpha$ precipitate than to its internal driving force. In any cases, the result is the stable growth of $\alpha|\gamma_{1-2}$ lamellae and it definitely showcases the sufficient role of kinetics (i.e. fast grain-boundary diffusion) under the two-local-equilibrium hypothesis on the origin of U-Nb's DP. Apparently, for the DP to happen within this phenomenological group, it is required the roles of both thermodynamics and kinetics, the importance is attached more to the later than former.

To remark, it is interesting to see from the above analyses how elastic strain affects in a non-trivial way the fundamental thermodynamics of the DP reaction. It is also interesting to see the reaction interface as a phase subjected to transformation and that such small changes in the lattice parameters of the reactants can alter the interface transformation path leading to a rich and rather-less-expected set of thermodynamic properties that explain experimental observations. And, although the current theoretical analysis is by no means comprehensive, the authors expect that local strain's complicated relation to misfit lattices, its importance, and its impact on DP hold true in general.

The differences in the observations of Djuric and Hackenberg \etal may be affiliated to different heat treatment conditions. In the case of Djuric, material processing perhaps yields a more relaxed sample with lattice parameters around grain boundary closer to equilibrium. It follows from here that the $\alpha$ precipitates nucleating from grain boundaries should have lattice parameters close to their equilibrium; small energy barrier for forming habit planes with $3X$ misfit unit should also be expected. The resulting elasto-chemical energies should be as such similar to those within the second group; and the resulting microstructure evolution contains the intermediate $\gamma$ precipitate with equilibrium composition evolves towards the higher value when temperature increases. Hackenberg \etal, on the other hand, could have samples with higher local strain at around grain boundaries as the consequence of more abrupt cooling. Consequently, when $\alpha$ precipitates form at the boundaries, there is a good chance that they possess larger lattice parameters. The elasto-chemical energies in this case will likely be similar to those belong to the first group. With their characteristic energy valley at around the equiatomic composition, the observed intermediate $\gamma$ precipitates would have their compositions close to equiatomic. 

It should be noted that the effect of the local strain at the grain boundaries does not affect final equilibrium of the system in both cases. The reason is simply that the stable energy valley would either resume after the relaxation of local strain for a prolonged heat treatment (i.e. discontinuous coarsening \cite{2011Hackenberg}) or already exists throughout the entire reaction (as demonstrated in many of the elasto-chemical potentials). It only tends to affect the DP reaction and is once again a good example of how processing affects the microstructures of materials and in turn changes the materials' performances.

\section{Conclusion}

To conclude, we have attempted to understand the discontinuous precipitation in U-Nb via thermodynamic and kinetic points of view. It has been suggested via our theoretical investigations that local strain can play an important role on the occurrence and stable growth of U-Nb's discontinuous precipitation. In particular, the strain helps to stabilize the metastable $\gamma$ phase by forming local energy valleys along the monotectoid decomposition path from initial $\gamma$ matrix to stable $\gamma_{2}$. Such energy valleys act as local traps that tend to arrest the reaction within the discontinuous-precipitation regime. The mechanism of the arrest can be strictly (local) equilibrium thermodynamics as similar to eutectoid reaction or both thermodynamics and kinetics. Which mechanism arresting the reaction within the discontinuous-precipitation regime decides the outcomes of microstructural chemistry (i.e. equiatomic or changing intermediate $\gamma$ composition), morphology, and ultimately the material's properties and performances. It will be dependent on how extensive the local strain at the grain boundary is and could be altered via different processing conditions. Via this work, the authors hope to contribute towards a better understanding of fundamental thermodynamics and kinetics that govern the discontinuous precipitation in the U-Nb system. This would be fundamental to the potential tailoring of fuel properties and performances via adjustments of processing conditions - one for instance would turn the discontinuous precipitation's disadvantages into its out-performance  \cite{2003Talach}.

\section*{Data availability }
The data that support the findings of this study are openly available upon request if available.

\section*{Acknowledgements}

This work was performed under the auspices of the United States Department of Energy by the Lawrence Livermore National Laboratory and Los Alamos National Laboratory under contract Nos. DE-AC52-07NA27344 and DE-AC52-06NA25396, respectively. Thien C. Duong specially thanks Prof. Ingo Steinbach, Dr. Oleg Shchyglo, Dr. Reza Darvishi Kamachali, Matthias Stratmann, Adam A. Gie{\ss}mann, and Efim Borukhovic for helpful discussions regarding phase-field theory and the interface dissipation model. Vahid Attari thanks the support by the National Science Foundation under NSF Grant No. CMMI-1462255. First-principles calculations were carried out in the Chemical Engineering Cluster and the Texas A\&M Supercomputing Facility at Texas A\&M University as well as in the Ranger Cluster at the Texas Advanced Computing Center at University of Texas, Austin.

\vspace{1cm}

\noindent \textbf{References}

\vspace{0.25cm}

\bibliographystyle{elsarticle-num}
\bibliography{references}

\begin{thebibliography}{10}
\expandafter\ifx\csname url\endcsname\relax
  \def\url#1{\texttt{#1}}\fi
\expandafter\ifx\csname urlprefix\endcsname\relax\def\urlprefix{URL }\fi
\expandafter\ifx\csname href\endcsname\relax
  \def\href#1#2{#2} \def\path#1{#1}\fi

\bibitem{1980Vandermeer}
R.~Vandermeer, Phase transformations in a uranium-14 at.\% niobium alloy, Acta
  Metallurgica 28~(3) (1980) 383--393.

\bibitem{1984Eckelmeyer}
K.~Eckelmeyer, A.~Romig, L.~Weirick, The effect of quench rate on the
  microstructure, mechanical properties, and corrosion behavior of
  \uppercase{U}-6 wt pct \uppercase{N}b, Metallurgical Transactions A 15~(7)
  (1984) 1319--1330.

\bibitem{2007Volz}
H.~M. Volz, R.~E. Hackenberg, A.~M. Kelly, W.~Hults, A.~Lawson, R.~Field,
  D.~Teter, D.~Thoma, X-ray diffraction analyses of aged
  \uppercase{U}-\uppercase{N}b alloys, Journal of Alloys and Compounds 444-445
  (2007) 217--225.

\bibitem{2009Clarke}
A.~Clarke, R.~Field, R.~Hackenberg, D.~Thoma, D.~Brown, D.~Teter, M.~Miller,
  K.~Russell, D.~Edmonds, G.~Beverini, Low temperature age-hardening in
  \uppercase{U}-13 at.\% \uppercase{N}b: an assessment of chemical
  redistribution mechanisms, Journal of Nuclear Materials 393 (2009) 282--291.

\bibitem{1981Williams}
D.~Williams, E.~Butler, Grain boundary discontinuous precipitation reactions,
  International Metals Reviews 26~(1) (1981) 153--183.

\bibitem{2001Manna}
I.~Manna, S.~Pabi, W.~Gust, Discontinuous reactions in solids, International
  Materials Reviews 46~(2) (2001) 53--91.

\bibitem{2011Hackenberg}
R.~E. Hackenberg, H.~M. Volz, P.~A. Papin, A.~M. Kelly, R.~T. Forsyth, T.~J.
  Tucker, K.~D. Clarke, Kinetics of lamellar decomposition reactions in
  \uppercase{U}-\uppercase{N}b alloys, Solid State Phenomena 172 (2011)
  555--560.

\bibitem{1972Djuric}
B.~Djuri{\'c}, Decomposition of gamma phase in a uranium-9.5 wt\% niobium
  alloy, Journal of Nuclear Materials 44~(2) (1972) 207--214.

\bibitem{2015Hackenberg}
{Hackenberg RE, Yablinsky CA, Llobet A, Volz HM, Papin PA, Tucker TJ, Clarke
  KD, Emigh MG.}, Lamellar and nonlamellar decomposition in {U-Nb}: Energy
  sinks and approach to equilibrium, in: Proc. of the Int. Conf. On Solid-Solid
  Phase Transformations in Inorganic Materials, M. Millitzer, et al. -ed. TMS
  Warrendale, PA, PTM, 2015, pp. 211--218.

\bibitem{1998Koike}
J.~Koike, M.~Kassner, R.~Tate, R.~Rosen, The \uppercase{N}b-\uppercase{U}
  (niobium-uranium) system, Journal of Phase Equilibria 19~(3) (1998) 253--260.

\bibitem{2008Liu}
X.~Liu, Z.~Li, J.~Wang, C.~Wang, Thermodynamic modeling of the
  \uppercase{U}-\uppercase{M}n and \uppercase{U}-\uppercase{N}b systems,
  Journal of Nuclear Materials 380~(1) (2008) 99--104.

\bibitem{2016Duong}
T.~C. Duong, R.~E. Hackenberg, A.~Landa, P.~Honarmandi, A.~Talapatra, H.~M.
  Volz, A.~Llobet, A.~I. Smith, G.~King, S.~Bajaj, et~al., Revisiting
  thermodynamics and kinetic diffusivities of uranium--niobium with bayesian
  uncertainty analysis, Calphad 55 (2016) 219--230.

\bibitem{2002Chen}
L.-Q. Chen, Phase-field models for microstructure evolution, Annual review of
  materials research 32~(1) (2002) 113--140.

\bibitem{2008Moelans}
N.~Moelans, B.~Blanpain, P.~Wollants, An introduction to phase-field modeling
  of microstructure evolution, Calphad 32~(2) (2008) 268--294.

\bibitem{2009Steinbach}
I.~Steinbach, Phase-field models in materials science, Modelling and Simulation
  in Materials Science and Engineering 17~(7) (2009) 073001.

\bibitem{attari2016phase}
V.~Attari, R.~Arroyave, Phase field modeling of joint formation during
  isothermal solidification in 3dic micro packaging, Journal of Phase
  Equilibria and Diffusion 37~(4) (2016) 469--480.

\bibitem{attari2018interfacial}
V.~Attari, S.~Ghosh, T.~Duong, R.~Arroyave, On the interfacial phase growth and
  vacancy evolution during accelerated electromigration in cu/sn/cu
  microjoints, Acta Materialia 160 (2018) 185--198.

\bibitem{2012Steinbach}
I.~Steinbach, L.~Zhang, M.~Plapp, Phase-field model with finite interface
  dissipation, Acta Materialia 60~(6) (2012) 2689--2701.

\bibitem{2012Zhang}
L.~Zhang, I.~Steinbach, Phase-field model with finite interface dissipation:
  extension to multi-component multi-phase alloys, Acta Materialia 60~(6)
  (2012) 2702--2710.

\bibitem{2009Amirouche}
L.~Amirouche, M.~Plapp, Phase-field modeling of the discontinuous precipitation
  reaction, Acta Materialia 57~(1) (2009) 237--247.

\bibitem{2011Amirouche}
L.~Amirouche, M.~Plapp, On the effect of bulk diffusion on the initiation of
  the discontinuous precipitation reaction: phase-field simulations, in: Solid
  State Phenomena, Vol. 172, Trans Tech Publ, 2011, pp. 549--554.

\bibitem{1955Turnbull}
D.~Turnbull, Theory of cellular precipitation, Acta Metallurgica 3~(1) (1955)
  55--63.

\bibitem{1959Cahn}
J.~W. Cahn, The kinetics of cellular segregation reactions, Acta Metallurgica
  7~(1) (1959) 18--28.

\bibitem{1972Fournelle}
R.~Fournelle, J.~Clark, The genesis of the cellular precipitation reaction,
  Metallurgical Transactions 3~(11) (1972) 2757--2767.

\bibitem{1991Fournelle}
R.~Fournelle, On the thermodynamic driving force for diffusion-induced grain
  boundary migration, discontinuous precipitation and liquid film migration in
  binary alloys, Materials Science and Engineering: A 138~(1) (1991) 133--145.

\bibitem{1972Hillert}
M.~Hillert, On theories of growth during discontinuous precipitation,
  Metallurgical Transactions 3~(11) (1972) 2729--2741.

\bibitem{1982Hillert}
M.~Hillert, An improved model for discontinuous precipitation, Acta
  Metallurgica 30~(8) (1982) 1689--1696.

\bibitem{1973Sundquist}
B.~E. Sundquist, Cellular precipitation, Metallurgical Transactions 4~(8)
  (1973) 1919--1934.

\bibitem{1997Klinger}
L.~Klinger, Y.~Brechet, G.~Purdy, On velocity and spacing selection in
  discontinuous precipitation-{I}. simplified analytical approach, Acta
  materialia 45~(12) (1997) 5005--5013.

\bibitem{2001Purdy}
G.~R. Purdy, Interface migration in diffusional phase transformations:
  Thermodynamic and kinetic aspects, in: Defect and Diffusion Forum, Vol. 194,
  Trans Tech Publ, 2001, pp. 1745--1758.

\bibitem{2013Robson}
J.~Robson, Modeling competitive continuous and discontinuous precipitation,
  Acta Materialia 61~(20) (2013) 7781--7790.

\bibitem{1960Peterson}
N.~L. Peterson, R.~E. Ogilvie, Diffusion studies in the uranium-niobium
  (columbium) system, Trans. Met. Soc. AIME 218 (1960) 439--444.

\bibitem{1963Petersonb}
N.~Peterson, R.~Ogilvie, Diffusion in the uranium-niobium (columbium) system,
  Trans. AIME 227 (1963) 1083--1087.

\bibitem{1963Eeles}
W.~Eeles, A.~Sutton, X-ray determination of the atomic positions in
  alpha-uranium at 22 and 660 degree c, Acta Crystallographica 16~(6) (1963)
  575.

\bibitem{2005Balluffi}
R.~W. Balluffi, S.~Allen, W.~C. Carter, Kinetics of materials, John Wiley \&
  Sons, 2005.

\bibitem{cahn1961spinodal}
J.~W. Cahn, On spinodal decomposition, Acta metallurgica 9~(9) (1961) 795--801.

\bibitem{yi2018strain}
S.-i. Yi, V.~Attari, M.~Jeong, J.~Jian, S.~Xue, H.~Wang, R.~Arroyave, C.~Yu,
  Strain-induced suppression of the miscibility gap in nanostructured
  {Mg2Si--Mg2Sn} solid solutions, Journal of Materials Chemistry A 6~(36)
  (2018) 17559--17570.

\bibitem{attari2019exploration}
V.~Attari, A.~Cruzado, R.~Arroyave, Exploration of the microstructure space in
  {TiAlZrN} ultra-hard nanostructured coatings, Acta Materialia 174 (2019)
  459--476.

\bibitem{EMTO}
L.~Vitos, Computational quantum mechanics for materials engineers: the EMTO
  method and applications, Springer Science \& Business Media, 2007.

\bibitem{1975Couterne}
A.~Couterne, C.~Collot, C.~Guillaume, Etude de l'alliage mulberry [u-7, 5 nb-2,
  5 zr (\% ponderaux)]. diagramme de transformation en refroidissement
  continu-structures et proprietes mecaniques, Journal of Nuclear Materials
  56~(2) (1975) 169--194.

\bibitem{1982Dahmen}
U.~Dahmen, Orientation relationships in precipitation systems, Acta
  Metallurgica 30~(1) (1982) 63--73.

\bibitem{1970Jackson}
R.~J. Jackson, Reversible martensitic transformation between transition phases
  of uranium-base niobium alloys., Tech. Rep. RFP-1535, Dow Chemical Co.,
  Golden, Colo. Rocky Flats Div. (1970).

\bibitem{1956Bridge}
J.~Bridge, C.~Schwartz, D.~Vaughan, X-ray diffraction determination of the
  coefficients of expansion of alpha uranium, Trans. AIME 206 (1956)
  1282--1285.

\bibitem{2002Cverna}
F.~Cverna, et~al., ASM ready reference: thermal properties of metals, ASM
  International, 2002.

\bibitem{1976Jackson}
R.~Jackson, J.~{Burke}, Elastic, plastic, and strength properties of {U--Nb}
  and {U--Nb--Zr} alloys, in: Physical metallurgy of uranium alloys, Brook Hill
  Publishing Co., 1976, pp. 611--656.

\bibitem{2013Jain}
A.~Jain, S.~P. Ong, G.~Hautier, W.~Chen, W.~D. Richards, S.~Dacek, S.~Cholia,
  D.~Gunter, D.~Skinner, G.~Ceder, K.~a. Persson, {The Materials Project: A
  materials genome approach to accelerating materials innovation}, APL
  Materials 1~(1) (2013) 011002.

\bibitem{1959Korchynsky}
M.~Korchynsky, R.~Fountain, Precipitation phenomena in cobalt-tantalum alloys,
  Trans. Met. Soc. AIME 215 (1959) 1033--1093.

\bibitem{2003Talach}
M.~Ta{\l}ach-Duma{\'n}ska, P.~Zieba, A.~Paw{\l}owski, J.~Wojewoda, W.~Gust,
  Practical aspects of discontinuous precipitation and dissolution, Materials
  chemistry and physics 80~(2) (2003) 476--481.

\end{thebibliography}

\clearpage

\end{document}